\documentclass[12pt]{article}
\usepackage{jheppub}
\usepackage[colorinlistoftodos]{todonotes}
\usepackage{url}
\usepackage{comment}
\usepackage{color}
\usepackage{xcolor}
\usepackage{soul}
\usepackage{tikz}
\usetikzlibrary{positioning,arrows.meta}
\usepackage{tikz-cd} 
\usepackage{physics}
\usepackage{lscape} 
\usepackage{tabularx}
\usepackage{cancel}
\usepackage{fancybox}
\usepackage{soul}
\usepackage{framed}
\usepackage{makecell}
\usepackage{multirow}
\usepackage[framemethod=tikz]{mdframed}

\newcommand{\PE}{\mathrm{PE}}
\newcommand{\BV}{\mathrm{BV}}

\newcommand{\Ber}{\mathrm{Ber}}

\newcommand{\bC}{\mathbb{C}}

\newcommand{\fgl}{\mathfrak{gl}}

\newcommand\cM{\mathcal M}

\newcommand\cZ{\mathcal Z}
\newcommand\cQ{\mathcal Q}
\newcommand\cD{\mathcal D}
\newcommand\cL{\mathcal L}
\newcommand\cA{\mathcal A}
\newcommand\cR{\mathcal R}
\newcommand\cH{\mathcal H}
\newcommand\sTr{\mathrm{sTr}}

\title{Geometric Quantization of the Giant Graviton Master Moduli Space}

\author[1]{Kasia Budzik,}
\author[2]{Davide Gaiotto}
\affiliation[1]{Center for the Fundamental Laws of Nature, Harvard University, Cambridge, MA 02138, USA}
\affiliation[2]{Perimeter Institute for Theoretical Physics, 31 Caroline Street North, Waterloo, ON N2L 2Y5, Canada}
\emailAdd{kbudzik@fas.harvard.edu}
\emailAdd{dgaiotto@perimeterinstitute.ca}

\abstract{We describe a remarkable reorganization of the space of gauge invariant operators of free or twisted $U(N)$ gauge theories as combinations of single-trace and determinant operators modulo certain ``open loop equation''. The reorganization has a transparent holographic interpretation, describing how states at finite $N$ can be reproduced by imposing certain $N$-independent gauge constraints on the combination of the closed strings Fock space and a geometric quantization of the moduli space of Giant Graviton D-branes, with quantization parameter $N^{-1}$. The moduli space of Giant Graviton D-branes is reproduced from tachyon condensation of auxiliary D-brane/anti-D-brane pairs which are ``Koszul dual'' to the D-brane system which supports the original gauge theory. We describe how to impose the open loop equations via certain BRST complexes and conjecture that their homology recovers the finite-$N$ BPS Hilbert space in the limit of large auxiliary rank. The BRST construction conjecturally categorifies Murthy’s Giant Graviton Expansion of indices counting $U(N)$-invariant BPS operators.}

\begin{document}
\maketitle

\section{Introduction and Conclusions}

The 't Hooft large $N$ expansion of $U(N)$ gauge theories is expected to match the perturbative expansion of dual String Theories 
\cite{tHooft:1973alw,Gross:1993hu,Polyakov:1987ez,Maldacena:1997re,Gubser:1998bc,Witten:1998qj,Aharony:1999ti,Bershadsky:1998mb,Gopakumar:1998ki,Ooguri:2002gx}. Finite $N$ gauge theories can thus be interpreted as a non-perturbative definition of these String Theories. Through holography \cite{Maldacena:1997re,Gubser:1998bc,Witten:1998qj}, this observation has been a source of non-perturbative insights on Quantum Gravity. 
In particular, the structure of the Hilbert space as a function of $N$ plays 
an important role in holographic studies of black holes and the information paradox \cite{Leutheusser:2021qhd,Leutheusser:2021frk,Witten:2021unn,Chandrasekaran:2022eqq,Leutheusser:2022bgi,Gesteau:2024rpt, Papadodimas:2012aq,Penington:2019npb,Marolf:2020xie,Schlenker:2022dyo}.

A general holographic expectation is that sectors of the Hilbert space with fixed quantum numbers will stabilize as $N$ is increased at fixed values of the 't Hooft couplings, converging to the Fock space of weakly-coupled closed strings \cite{Sundborg:1999ue,Aharony:2003sx,Kinney:2005ej,Chang:2013fba}. For example, in a conformal gauge theory we can identify the space of states on a sphere as the space of local operators, which stabilizes to polynomials in single-trace operators dual to individual closed string states \cite{Gubser:1998bc,Witten:1998qj,Berenstein:2002jq}. 

This description breaks down for states with quantum numbers of order $N$ \cite{Maldacena:1998bw,McGreevy:2000cw,Murthy:2020scj}. For example, the trace relations satisfied by single-trace operators \cite{Razmyslov1974,Procesi1976} may remove states from the spectrum \cite{Benvenuti:2006qr,Dolan:2007rq}, while in interacting BPS sectors they can also give rise to new BPS states \cite{Chang:2022mjp,Choi:2022caq,Choi:2023znd,Chang:2024zqi}. There is a general expectation that these features of gauge theory should be explained holographically by the contribution of ``Giant Graviton'' D-branes  \cite{McGreevy:2000cw,Grisaru:2000zn,Hashimoto:2000zp,Mikhailov:2000ya,Balasubramanian:2001nh,Corley:2001zk,Berenstein:2004kk} to the spectrum, but the details are unclear. In particular, the D-brane contribution must somehow subtract certain states from the gauge theory Fock space to match the effect of trace relations. The so-called ``Giant Graviton expansions'' express finite $N$ indices as sums over D-brane contributions \cite{Arai:2020qaj,Imamura:2021ytr,Gaiotto:2021xce,Murthy:2022ien,Imamura:2022aua,Lee:2022vig,Liu:2022olj,Eleftheriou:2023jxr} (see \cite{Arai:2019xmp,Arai:2019aou,Arai:2020uwd} for earlier computations of the leading corrections). A recent proposal is that Giant Graviton D-branes should be given an alternative quantization which makes them into ``ghost'' states, which implement the trace relations on the closed string Fock space in a BRST fashion \cite{Lee:2023iil,Lee:2024hef}. 

A related puzzle is that a semi-classical quantization of Giant Graviton D-branes can sometimes account even for states with finite quantum numbers, with individual single-trace operators reproduced as the contribution from ``small'' D-branes \cite{Biswas:2006tj,Mandal:2006tk,Martelli:2006vh,Bhattacharyya:2007sa,Ashok:2008fa,Pasukonis:2012zj}. These results strongly suggests that individual gauge theory states may have multiple holographic descriptions \cite{Berenstein:2004kk}.

In this paper we propose the following resolution of these puzzles: the semi-classical quantization of Giant Graviton D-branes should be amended to include the effect of gauge symmetries which can convert open strings to closed strings. These gauge symmetries prevent the overcounting of states which could arise either from closed strings or small D-branes. Trace relations are still BRST-exact in this formulation, but the cancellations involve a disparate collection of closed strings and Giant Gravitons of different sizes.

The main focus of our analysis will be operators and states which preserve some supersymmetry (BPS). Recall that supersymmetric configurations can be often isolated from the rest of the theory and Witten indices of the spaces of local operators or states can be computed via localization techniques \cite{Witten:1982df,Romelsberger:2005eg,Kinney:2005ej,Bhattacharya:2008zy,Assel:2014paa,Rastelli:2016tbz}. Homological algebra techniques \cite{Grant:2008sk,Chang:2013fba,Chang:2022mjp,Choi:2022caq,Budzik:2023xbr} give access to the actual spaces of BPS states and can be naturally interpreted in the language of twisted supersymmetric gauge theory \cite{Johansen:1994aw,Costello:2013zra,Eager:2018oww,Eager:2018dsx,Budzik:2023xbr}. 

Our analysis is entirely algebraic and independent of the details of a specific theory. Indeed, it can be applied to a very general class of large $N$ counting problems. The whole analysis takes place in the large $N$ gauge theory: terms such as ``closed strings'', ``D-branes'', ``Giant Gravitons'' etc. are used to denote gauge theory structures which have the appropriate interpretation in a 't Hooft expansion. 

In many examples, the BPS subsectors have holographic descriptions involving topological string theories which capture BPS configurations of conventional theories of superstrings \cite{Costello:2016mgj,Costello:2018zrm,Ishtiaque:2018str}. In that context, we can map our construction to concrete D-branes \cite{Budzik:2021fyh}. A lift to conventional string theory duals may also be possible but likely more challenging. We leave such a lift, the analysis of loop corrections \cite{Beisert:2003jj,Budzik:2022mpd,Budzik:2023xbr,Choi:2025bhi, Budzik:2025zvu} and extensions to nonsupersymmetric sectors to future work. 

Quantum Gravity phenomena such as the appearance of black holes are expected to control the spectrum of states with quantum numbers of order $N^2$. This is expected to happen even in BPS sectors \cite{Gutowski:2004ez,Chong:2005hr,Kunduri:2006ek,Hosseini:2017mds,Berkooz:2006wc,Cabo-Bizet:2018ehj,Choi:2018hmj,Benini:2018ywd,Murthy:2020scj}. An obvious question is how to disentangle such effects, if at all possible, from the contributions of large numbers of D-branes \cite{Myers:2001aq,Balasubramanian:2005mg,Choi:2022ovw,Lee:2022vig,Beccaria:2023hip}. 
We expect our strategy to quantize Giant Gravitons and closed strings to account for all states in the gauge theory, including these which are associated to Black Holes and other nontrivial gravitational saddles \cite{Lin:2004nb,Aharony:2021zkr}. 

There are many other examples where the small coupling limit of supersymmetric black-hole microstates has been explained in terms of bound states of D-branes \cite{Strominger:1996sh,Breckenridge:1996is,Maldacena:1997de}. 
Our results could be interpreted in the same spirit, as providing \emph{weak-coupling BPS black hole microstate counting via D-branes} in a very general setting.

\subsection{Determinants and Redundancies}
At first, we will focus on BPS operators in a free theory: $U(N)$-invariant polynomials of a collection of adjoint BPS letters. 
We denote the space of $U(N)$ invariants as $\cH_N$. This is a truncation of a space $\cH_\infty$ of polynomials in single trace operators $\Tr W_\alpha$ \cite{Procesi1976,Razmyslov1974}. Here $W_\alpha$ are adjoint ``words'', matrix products of a sequence of BPS letters. 

Giant Graviton D-branes are conventionally associated to determinant operators in the gauge theory \cite{Balasubramanian:2001nh,Corley:2001zk,Berenstein:2003ah,Jiang:2019xdz,Budzik:2021fyh},  extending the identification of baryonic operators with wrapped D-branes \cite{Witten:1998xy,Gubser:1998fp,Beasley:2002xv}.\footnote{Open strings ending on Giant Gravitons correspond to words inserted into determinant operators \cite{Balasubramanian:2002sa,Berenstein:2003ah,deMelloKoch:2007rqf,Hofman:2007xp} producing modified determinants and are included in the generating function below.} The most general determinant operator takes the form 
\begin{equation}
    \det_{N \times N} \left[m_0 + \sum_{\alpha \neq 0} m_\alpha W_\alpha\right] \, ,
\end{equation}
where $m_\alpha$ are a collection of parameters. In a model-independent context, we identify any such operator as a ``Giant Graviton" D-brane and the $m_\alpha$ as 
(projective) coordinates on a moduli space of D-branes \cite{Mikhailov:2000ya,Beasley:2002xv,Budzik:2021fyh}, described as the tachyon condensation \cite{Sen:1998sm,Douglas:2000gi} of elementary building blocks we dub ``Koszul-dual branes'' \cite{Budzik:2025vzr}.

At finite $N$, the determinant operator can be expanded in a basis of degree $N$ monomials in the $m_\alpha$. The result is some collection of local operators in the gauge theory. In specific examples, this expansion was interpreted as a geometric quantization of the moduli space \cite{Beasley:2002xv,Biswas:2006tj,Pasukonis:2012zj}. Recall that geometric quantization of a K\"ahler manifold builds the Hilbert space as the space of holomorphic sections of the $N$-th power of a line bundle $\cL$. For a projective space, these are precisely monomials of degree $N$. 

Consider also the generating function 
\begin{equation}
    \cZ_0(t_\bullet) \equiv \exp \sum_{[\alpha]} t_{[\alpha]} \Tr W_\alpha
\end{equation}
of polynomials in single-trace operators. The trace relations which appear at finite $N$ kill certain linear combination of monomials and thus constrain the $t_\bullet$ dependence, but the constraint is difficult to write explicitly. 

Something interesting happens if we combine single-trace operators and determinant operators into a bigger generating function 
\begin{equation}
    \cZ_1(t_\bullet,m_\bullet) \equiv \cZ_0(t_\bullet)\det_{N \times N} \left[m_0 + \sum_\alpha m_\alpha W_\alpha\right] \, .
\end{equation}
As any operator can be written as a polynomial in single-trace operators, this generating function is highly redundant. A consequence of this redundancy is that
$\cZ_1(t_\bullet,m_\bullet)$ satisfies a collection of $N$-independent ``open loop equations'' 
\begin{equation}
    \cD_{\pm,\alpha} \cZ_1(t_\bullet,m_\bullet) =\left(\partial_{t_{[\alpha]}} - m_0 \partial_{m_\alpha} + \cdots\right)\cZ_1(t_\bullet,m_\bullet) \, , \label{eq:loopeq}
\end{equation}
where $\cD_{\pm,\alpha}$ are certain differential operators in $t_\bullet$ and $m_\bullet$. 

Our key observation is that the open loop equations become very powerful when combined with the requirement that $\cZ_1(t_\bullet,m_\bullet)$ is homogeneous of degree $N$ in $m_\alpha$. Indeed, we find that they fully capture trace relations:
\begin{framed}
The space $\cH_N$ of $U(N)$-invariant operators is dual to the space of solutions  of the open loop equations \eqref{eq:loopeq} which are degree $N$ homogeneous polynomials in $m_\bullet$ with coefficients polynomial in $t_\bullet$.
\end{framed}

This characterization gives $\cH_N$ as the output of a gauged geometric quantization procedure, i.e. as holomorphic sections constrained by some differential equations expressing the gauge redundancy:
a redefinition 
\begin{equation} \label{eq:gaugedet}
    \left[m_0 + \sum_\alpha m_\alpha W_\alpha\right] \to \left[1 + \sum_\alpha \ell_\alpha W_\alpha\right]\left[m_0 + \sum_\alpha m_\alpha W_\alpha\right]\left[1 + \sum_\alpha r_\alpha W_\alpha\right]
\end{equation}
can be absorbed into a shift of the $t_{[\alpha]}$'s.

\subsection{Derived Moduli Space of Giant Gravitons}
Many BPS counting problems involve the cohomology of complexes of $U(N)$-invariant words acted upon by some nilpotent differential $Q$ \cite{Kinney:2005ej,Grant:2008sk,Budzik:2023xbr}. The cohomology $\cH_{Q;N}$ of BPS operators truncated to some finite quantum numbers typically stabilizes at large $N$ \cite{Chang:2013fba}, but the cohomology with quantum numbers of order $N$ can jump in intricate ways due to the interaction between $Q$ and trace relations \cite{Chang:2022mjp, Choi:2022caq, Choi:2023znd, Chang:2024zqi}. In this paper we restrict our considerations to 
differentials computed at tree level and leave loop corrections to future work. 

In order to extend our considerations to a cohomological setting, we need a formalism to study the action of $Q$ on determinant operators. It is also useful to recast the open loop equations in a cohomological way, i.e. to apply a BRST formalism to the gauge redundancies they express. At least for the leading action of $Q$ (in a power series in the coupling), both goals can be accomplished by the same construction.  

The key observation is that the open loop equations \eqref{eq:loopeq} are Schwinger-Dyson equations for an auxiliary Grassmann integral which computes the determinant
\begin{equation}
    \det_{N \times N} \left[m_0 + \sum_\alpha m_\alpha W_\alpha\right] = \int \dd\psi \dd \bar \psi \, \exp\left[m_0 +  \sum_{\alpha} m_{\alpha} \bar \psi W_\alpha \psi\right] \, ,
\end{equation}
where $\bar\psi,\psi$ are (anti)fundamental zero-dimensional fermions.

The BV formalism is precisely designed to impose the Schwinger-Dyson equations cohomologically. One introduces antifields $u$, $\bar u$ for the fermions and a BV differential $\Delta_\BV$ such that
passing to $\Delta_\BV$ cohomology is equivalent to doing the fermionic integral. Accordingly, one could safely replace the 
action of $Q$ on BPS letters with the action of $Q + \Delta_\BV$ on 
polynomials of BPS letters, fermions and their antifields. 

We can then attempt to replace the action on individual letters with an uniform-in-$N$ action on $U(N)$-invariants: formal variables $T_\alpha$ and $\cM_\alpha$ representing traces $\Tr W_\alpha$ and 
mesons $\bar \psi W_\alpha \psi$, $\bar u W_\alpha \psi$, $\bar \psi W_\alpha u$, $\bar u W_\alpha u$. The good news is that this formal procedure automatically handles finite-$N$ trace relations, because they are consequences of integration-by-part identities. 
The bad news is that working uniformly in $N$ breaks some good properties of the  $\Delta_\BV$ cohomology, so that the cohomology of $Q + \Delta_\BV$ only gives an approximation of $\cH_{Q;N}$. We find experimentally that the cohomology map into $\cH_{Q;N}$ is not injective in some examples. It may also fail to be surjective, although we have not found explicit examples of that.

The next step of our analysis is to consider ``$k$-determinant'' operators involving $k$ pairs of epsilon tensors for $U(N)$. We obtain matrix-valued open loop equations which again potentially carve out $\cH_{Q;N}^\vee$ out of a large space of $k$-determinants dressed by single-trace operators, and an uniform-in-$N$ cohomology problem 
involving super-matrices $\cM_\bullet$ of mesons of dimension $k|k$. 
Experimentally, we observe that the challenges we encounter at $k=1$ are ameliorated as $k$ increases, giving an ever improving approximation to $\cH_{Q;N}$. We conjecture that $\cH_{Q;N}$ is recovered in a large $k$ limit:
\begin{mdframed}[
    linecolor=none,
    tikzsetting={draw=black,dashed,line width=0.4pt},
    innertopmargin=6pt,
    innerbottommargin=15pt
]\begin{align}
    \lim_{k\rightarrow\infty} H_\bullet(\mathfrak{gl}_{k|k}'[\cA^!]|\mathfrak{b}_{k|k}, \Ber^N) \simeq \cH_{Q;N} \, , \label{eq:again}
\end{align}\end{mdframed}
where the left hand side is a specific cohomology problem we formulate in the main text of the paper, involving the action of $Q + \Delta_\BV$ on polynomials of $T_\bullet$ and $\cM_\bullet$. 

\subsection{Koszul Duality and Murthy's Giant Graviton Expansion}

The construction of the previous subsection has a natural string theory interpretation. The algebra of words $\mathcal A^!$ is the open string algebra of auxiliary D-branes which, in twisted holography examples, are fully transverse to the original stack of $N$ D-branes, aka \emph{Koszul-dual} branes. Giant Graviton branes are described by tachyon condensation \cite{Douglas:2000gi,Vafa:2001qf} on pairs of these branes and antibranes.

Our BRST formalism therefore computes the geometric quantization of a moduli space of D-branes, combined with a Fock space of closed strings and quotiented by the gauge symmetries of the D-branes in a manner which accounts the possible creation of closed strings by the gauge transformations. We thus claim that 
\begin{framed}
    The finite $N$ BPS cohomology $\cH_{Q;N}$ of local operators for the (leading) action of the supercharge $Q$ can be reproduced by the geometric quantization of Giant Graviton D-branes with closed strings. 
\end{framed}
For example, in the cohomology problem for $1/16$-BPS states in ${\cal N}=4$ SYM, the Koszul-dual branes are D5-branes in $\mathbb{C}^5$, transverse to the stack of D3-branes which support the holomorphic twist of ${\cal N}=4$ SYM. Tachyon condensation on these D5-branes and antibranes produces B-model Giant Graviton D3-branes. Therefore our construction can be identified with the geometric quantization of a moduli space of Giant Graviton D3-branes. This is the expected twisted version of the conventional AdS${}_5\times$S${}^5$ IIB holographic dual \cite{Costello:2016mgj} and Giant Graviton D-branes \cite{Mikhailov:2000ya,Biswas:2006tj,Budzik:2021fyh}.

The idea that D-brane contributions may account for BPS states/operators of arbitrary charge originates from the Giant Graviton Expansions \cite{Gaiotto:2021xce,Imamura:2021ytr,Murthy:2022ien,Eleftheriou:2023jxr,Chen:2024cvf,Eniceicu:2023uvd,Beccaria:2024szi,Hayashi:2024aaf,Beccaria:2024oif,Fujiwara:2023bdc,Liu:2022olj,Ezroura:2024wmp,Lee:2022vig,Beccaria:2024dxi}: general combinatorial formulae which recast the Witten indices 
\begin{equation}
    Z_N \equiv \chi(\cH_N)
\end{equation}
as a sum of systematic corrections to the large $N$ limit $Z_\infty$. Individual corrections are interpreted as Giant Graviton contributions, though the details depend on the specific formula. 

Murthy's GGE \cite{Murthy:2022ien} applies to a broad class of partition functions $Z_N$ and takes the form  
\begin{equation}
    Z_N = Z_\infty\sum_{k=0}^\infty G_{N}^{(k)} \, . \label{eq:MGGE}
\end{equation}
Here $Z_N$ is computed by the standard $U(N)$ Molien-Weyl integral and $Z_\infty$ is its large $N$ limit. The corrections $G^{(k)}_N$ resemble $U(k|k)$ supergroup integrals with an extra insertion of the Berezinian to the $N$th power \cite{Liu:2022olj,Eniceicu:2023uvd,Chen:2024cvf}.

Building on the formula for supergroup invariants proposed by one of the authors \cite{Budzik:2025vzr}, we conjecture an extension to relative cohomology with coefficients in $\Ber^N$. Applied to our complex, this gives the conjectural identity
\begin{equation}
    \chi\left( H_\bullet(\mathfrak{gl}_{k|k}'[\cA^!]|\mathfrak{b}_{k|k}, \Ber^N)\right)= Z_\infty \sum_{l=0}^k G^{(l)}_N \, , 
\end{equation}
which approaches $Z_N$ as $k\rightarrow\infty$. Therefore our conjecture \eqref{eq:again} provides a \emph{categorification} of the Murthy's Giant Graviton expansion.

\subsection{Structure of the Paper}

In Section \ref{sec:counting}, we introduce the Open Loop Equations for the case of $d$ bosonic letters and generalize them to $k$-determinant operators and fermionic letters. In Section \ref{sec:ggmoduli}, we develop the dual description in terms of a moduli space of Giant Gravitons, interpreting the Open Loop Equations as gauge constraints. In Section \ref{sec:QOLE}, we build a BRST complex implementing these constraints cohomologically. In Section \ref{sec:MGGE}, we compare its graded characters with the Murthy's Giant Graviton expansion based on which we conjecture\footnote{Throughout the Sections below, we denote by solid boxes proven statements and by dashed boxes conjectures. The conjectures have been tested by using Python code produced by ChatGPT and Claude, which we plan to make available in version 2.} that the BRST complex will reproduce $\cH_{N}$ as $k \to \infty$.

%%%%%%%%%%%%%%%%%%%%%%%%%%%%%%%%%
\section{Counting and Re-counting}
\label{sec:counting}

The objective of this Section is to propose alternative descriptions of the space of $U(N)$-invariants, which trade trace relations for gauge-like constraints in an enlarged system. The enlarged system has the formal structure of a Fock space of closed strings combined with the geometric quantization of a D-brane phase space, quotiented by a common gauge symmetry.

For clarity of presentation, we focus here on the spaces $\mathcal{H}_{d;N}$ of $U(N)$-invariant polynomials built from $d$ bosonic $N \times N$ matrices $X_i$, transforming in the adjoint representation of $U(N)$. We will include fermionic matrices at a later stage. It is straightforward to modify our construction to include (anti)fundamental variables as well. It would be interesting to extend it to more general classical gauge groups.

The space $\mathcal{H}_{1;N}$ is well understood: the Cayley–Hamilton theorem allows one to express all $U(N)$-invariants as polynomials in traces $\Tr X^n$ with $n\leq N$, which do not satisfy any further relations, i.e.
\begin{equation}
    \mathcal{H}_{1;N} = \bC[\Tr X, \cdots, \Tr X^N] \, .
\end{equation}
We lack a similar characterization for $d>1$. The main conceptual challenge is the complexity of the polynomial ``trace relations'' satisfied by the traces of products of multiple matrices. See \cite{deMelloKoch:2025ngs, deMelloKoch:2025cec, deMelloKoch:2025qeq, deMelloKoch:2025rkw, Caputa:2025ikn, deMelloKoch:2025eqt, deMelloKoch:2026dfo} for recent interesting conjectures concerning the space of invariants of multiple matrices (and vectors) at finite $N$.

We will now introduce some convenient notation to discuss $U(N)$-invariant operators built from matrices $X_i$, $i=1,\dots,d$. We denote as $W_\alpha$ the product of any sequence of matrices (a word), e.g.
\begin{equation}
    W_{1232} \equiv X_1 X_2 X_3 X_2 \, .
\end{equation}
We have an obvious concatenation product $W_{\alpha} W_{\beta} = W_{\alpha \beta}$ defining the tensor algebra of $\bC^d$. We only consider non-trivial words unless otherwise stated, i.e. whenever we sum over $\alpha$ we do not include $\alpha = \emptyset$.

The traces
\begin{equation}
    \Tr W_\alpha
\end{equation}
depend only on a cyclic ordering $[\alpha]$ of the matrices, e.g. 
\begin{equation}
    \Tr X_1 X_2 X_3 X_2 = \Tr  X_2 X_1 X_2 X_3 \, .
\end{equation}
A general combinatorial result is that all $U(N)$-invariant polynomials in the $X_i$ can be written, possibly in multiple ways, as polynomials in the $\Tr W_\alpha$. The redundancy is due to trace relations, which only affect expressions involving more than $N$ matrices. 

A convenient way to encode this statement is to introduce the space $\mathcal{H}_{d;\infty} = \bC[T_\bullet]$ of polynomials in a collection of variables $T_{[\alpha]}$ representing the traces at $N=\infty$. We can then define a surjective map
\begin{equation}
   \pi_N: \mathcal{H}_{d;\infty} \rightarrow \mathcal{H}_{d;N} \, , \qquad\qquad  \pi_N\qty(T_{[\alpha]})=\Tr W_\alpha \, .
\end{equation}
We will refer to the kernel of this map as the space ${\cal T}_{d;N}$ of \emph{trace relations}, i.e. we have
\begin{equation}\label{eq:tracerel}
    0 \to {\cal T}_{d;N} \to \mathcal{H}_{d;\infty} \to \mathcal{H}_{d;N} \to 0 \, . 
\end{equation}

\subsection{Determinant Modifications}

A well-known result \cite{Procesi:1976} is that all trace relations follow from a simple statement: the exterior products $\Lambda^{n} \bC^N$ are trivial for $n>N$ and accordingly
\begin{equation}
    \Tr_{\Lambda^{n} \bC^N} (W_{\alpha_1}\wedge \cdots \wedge W_{\alpha_n}) \equiv \frac{1}{n!}\sum_{\sigma\in S_n} (-1)^{|\sigma|} \Tr\, (W_{\alpha_1})^{i_1}_{i_{\sigma(1)}} \dots (W_{\alpha_n})^{i_n}_{i_{\sigma(n)}}
\end{equation}
vanish for $n>N$.

We can define polynomials $\Lambda_{(\alpha_1, \cdots, \alpha_{n})}\in \bC[T_\bullet]$, whose image in $\mathcal{H}_{d;N}$ coincides with $n! \Tr_{\Lambda^{n} \bC^N} W_{\alpha_1}\wedge \cdots \wedge W_{\alpha_n}$.
For example,
\begin{equation}
	\Lambda_{(\alpha)}= T_{[\alpha]}
\end{equation}
maps to $\Tr W_\alpha$ and is a trace relation for $N=0$, 
\begin{equation}
	 \Lambda_{(\alpha,\beta)}= T_{[\alpha]}T_{[\beta]} - T_{[\alpha \beta]} 
\end{equation}
maps to 
\begin{equation}
    2 \Tr_{\Lambda^{2} \bC^N} W_{\alpha_1}\wedge W_{\alpha_2} = \Tr W_\alpha \Tr W_\beta- \Tr W_\alpha W_\beta
\end{equation}
and is a trace relation for $N=0,1$, \begin{align}	 \Lambda_{(\alpha,\beta,\gamma)}= T_{[\alpha]} T_{[\beta]} T_{[\gamma]} - T_{[\alpha]} T_{[\beta \gamma]} - T_{[\gamma]} T_{[\alpha \beta]} -  T_{[\beta]}  T_{[\gamma \alpha]} +  T_{[\alpha\beta\gamma]} + T_{[\alpha\gamma \beta]}
\end{align}
is a trace relation for $N=0,1,2$, etc. 

Note that $\Lambda_{(\alpha_1, \cdots, \alpha_{n})}$ are  symmetric in the arguments. We will often employ a multi-index notation $\Lambda_A$, with $A$ denoting a symmetric collection of $\alpha$'s and $|A|=n$.

The $\Lambda_A$ are often referred to as \emph{subdeterminant} operators, as they appear as coefficients in the universal expansion of determinant operators:
\begin{align}
    1+ \sum_A \frac{m_A}{|\mathrm{Aut} A|} \Lambda_A  \rightarrow \det \left[1 + \sum_{\alpha} m_\alpha W_\alpha \right] = \exp \Tr \log\left[1 + \sum_{\alpha} m_\alpha W_\alpha \right] \, ,
\end{align}
under $\pi_N:T_{[\alpha]} \rightarrow \Tr W_\alpha$.

\subsection{Open Loop Equations}
\label{subsec:OLE}

We will now describe a collection of ``open loop equations" which are linear relations on $\Lambda_\bullet$ with $T_\bullet$-dependent coefficients. These algebraic identities have appeared previously in
the math literature \cite{vaccarino, buchstaber2001gelfand}, and related Schwinger-Dyson equations appear in \cite{Gaiotto:2021xce, Choi:2026faq}. To the best of our knowledge, the explicit $\mathbb C[T_\bullet]$-module
presentation of $\mathcal{H}_{d;N}$ and $\mathcal{T}_{d;N}$ developed below is new.

We can reformulate the determinant as a Grassmann integral over $2N$ auxiliary fermions $\psi$ and $\bar \psi$:
\begin{equation}
     \det \left[1 + \sum_{\alpha} m_\alpha W_\alpha \right]  = \int \dd \psi \dd \bar \psi \, e^{\bar \psi \psi + \sum_{\alpha} m_\alpha \bar \psi W_\alpha \psi} \, . \label{eq:grass}
\end{equation}

This integral satisfies integration-by-part identities, i.e. Schwinger-Dyson equations:
\begin{equation}
\begin{split}
    &\int \dd \psi \dd \bar \psi \, \partial_\psi \left[  e^{\bar \psi \psi + \sum_{\alpha} m_\alpha \bar \psi W_\alpha \psi} W_\alpha \psi \right] =0 \cr
     &\int \dd \psi \dd \bar \psi \, \partial_{\bar \psi} \left[  e^{\bar \psi \psi + \sum_{\alpha} m_\alpha \bar \psi W_\alpha \psi} \bar \psi W_\alpha \right] =0 \, .
\end{split}
\end{equation}
Expanding in powers of $m_\alpha$, we find $N$-independent equations which are the image of relations
\begin{align}
	 \Lambda_{(\alpha, \beta_1, \cdots, \beta_n)} + \sum_{i=1}^n \Lambda_{(\beta_1, \cdots, \alpha \beta_i, \cdots,  \beta_n)}&=T_{[\alpha]} \Lambda_{(\beta_1, \cdots, \beta_n)} \,  \label{eq:ole1} \\
	\Lambda_{(\alpha, \beta_1, \cdots, \beta_n)} + \sum_{i=1}^n \Lambda_{(\beta_1, \cdots, \beta_i \alpha , \cdots,  \beta_n)} &= T_{[\alpha]} \Lambda_{(\beta_1, \cdots, \beta_n)} \,  \label{eq:ole2}
\end{align}
in $\bC[T_\bullet]$. We denote these relations as \textit{the open loop equations}. 

The open loop equations can also be understood as the infinitesimal version of the obvious determinant factorization identity: 
\begin{align}
    &\det_{N \times N}\!\left[1 + \sum_\alpha \ell_\alpha W_\alpha \right]\left[m_0 + \sum_\alpha m_\alpha W_\alpha \right]\left[1 + \sum_\alpha r_\alpha W_\alpha \right]  =\cr &= \exp\left( \Tr \log \left[1 + \sum_\alpha \ell_\alpha W_\alpha \right] \left[1 + \sum_\alpha r_\alpha W_\alpha \right]\right) \det_{N \times N}\!\left[m_0 + \sum_\alpha m_\alpha W_\alpha \right] \, . \label{eq:sym}
\end{align}

The open loop equations manifestly allow one to recursively reconstruct the $\Lambda_A$ as polynomials in $T_\bullet$. We will find it useful to define a $\bC[T_\bullet]$-module $\cR$ with generators $R_A$ (including $R_\emptyset$) and relations 
\begin{align}
\begin{split}
	 R_{(\alpha, \beta_1, \cdots, \beta_n)} + \sum_{i=1}^n R_{(\beta_1, \cdots, \alpha \beta_i, \cdots,  \beta_n)} &= T_{[\alpha]} R_{(\beta_1, \cdots, \beta_n)}\cr
     R_{(\alpha, \beta_1, \cdots, \beta_n)} + \sum_{i=1}^n R_{(\beta_1, \cdots, \beta_i \alpha , \cdots,  \beta_n)}&= T_{[\alpha]} R_{(\beta_1, \cdots, \beta_n)} \, 
\end{split} \label{eq:role}
\end{align}
and a module map $R_A \to \Lambda_A$ to $\bC[T_\bullet]$. As the relations allow to reduce all $R$'s to an element in $\bC[T_\bullet]R_\emptyset$, this is obviously an isomorphism:
\begin{equation}
    \cR \equiv \frac{\oplus_A \bC[T_\bullet]R_A}{\mathrm{OLE}} \simeq \bC[T_\bullet] \, .
\end{equation}
It suffices to impose the open loop equations for one representative of each cyclic class of $\alpha$, since the remaining equations follow from the commutator relations discussed in Section \ref{subsec:OLEalgebra}.

We will now describe both $\cH_{d;N}$ and ${\cal T}_{d;N}$ in the same language. As trace relations are the image of $R_A$'s with $|A|>N$, we find
\begin{framed}\begin{equation}
    \cH_{d;N} \simeq \cR_{\leq} \equiv \frac{\oplus_A \bC[T_\bullet]R_A}{\mathrm{OLE} \cup \oplus_{|A|>N} \bC[T_\bullet]R_A} \, .
\end{equation}\end{framed}
\noindent Concretely, the $\bC[T_\bullet]$-module $\cR_{\leq}$ is generated by the $R_A$ with $|A|\leq N$, subject to the open loop equations \eqref{eq:role} with $n\leq N$ truncated by setting the remaining $R_A$ to $0$.

A basic theorem in linear algebra gives the short exact sequence 
\begin{equation}
    0\to \frac{\oplus_{|A|>N} \bC[T_\bullet]R_A}{\mathrm{OLE} \cap \oplus_{|A|>N} \bC[T_\bullet]R_A} \to \cR \to \cR_{\leq } \to 0 \, .
\end{equation}
This characterizes the space $\mathcal{T}_{d;N}$ of trace relations as 
the quotient of $\oplus_{|A|>N} \bC[T_\bullet]R_A$ by all linear combinations of open loop equations which involves these $R_A$'s only:
\begin{framed}\begin{equation}
    \mathcal{T}_{d;N} \simeq \frac{\oplus_{|A|>N} \bC[T_\bullet]R_A}{\mathrm{OLE} \cap \oplus_{|A|>N} \bC[T_\bullet]R_A} \, .
\end{equation}\end{framed}

We can define a simpler $\bC[T_\bullet]$-module $\cR_{>}$ by quotienting $\oplus_{|A|>N} \bC[T_\bullet]R_A$ only by the open loop equations involving $R_A$ with $|A|>N$. In principle, $\cR^{>}$ could be bigger than $\mathcal{T}_{d;N}$, so that the first map in the sequence
\begin{equation}
    \cR_{>}\to \cR \to \cR_{\leq } \to 0 \, .
\end{equation}
is not injective. We conjecture that
\begin{mdframed}[
    linecolor=none,
    tikzsetting={draw=black,dashed,line width=0.4pt},
    innertopmargin=12pt,
    innerbottommargin=15pt
]
\begin{equation}
    \mathcal{T}_{d;N} \simeq \mathcal{R}_{>} \equiv \frac{\oplus_{|A|>N} \bC[T_\bullet]R_A}{\mathrm{OLE}_{|A|>N}} \, .
\end{equation}
\end{mdframed}
\noindent Here, $\mathrm{OLE}_{|A|>N}$ denotes open loop equations involving only generators $R_A$ with $|A|>N$. Experimentally, $\mathcal{T}_{d;N}\simeq \cR_{>} $ for all examples we tested. This conjecture will not be necessary for the cohomological constructions described later in this paper. 

We can think about the $R_A$ generators as monomials in a collection of $M_\alpha$ auxiliary variables, representing the meson operators $\bar \psi W_\alpha \psi$ which can be inserted into an $\int \dd \psi \dd \bar \psi \, e^{\bar \psi \psi}$ Grassmann integral to produce the $\Lambda_A$:
\begin{align}
    R_{(\alpha_1,\dots,\alpha_n)} \equiv M_{\alpha_1} \dots M_{\alpha_n} \, .
\end{align}
Accordingly, we identify $\cH_{d;N}$ as the quotient of $\bC[T_\bullet, M_\bullet]$ by two constraints: the open loop equations and the requirement that the product of more than $N$ mesons vanishes. 

We will find sometimes useful to introduce an extra meson ``zeromode'' $M_0$, mapping to $\bar \psi \psi$, so that we can work with polynomials in $\bC[T_\bullet, M_\bullet,M_0]$ of degree exactly $N$ in the $M$'s and use the bare integration map $\int \dd \psi \dd \bar \psi$.

We can generalize these constructions to ``multi-determinants". These are computed by a Grassmann integral over $2 k N$ fermions
\begin{equation}
     \det_{kN\times kN} \left[\mathbf{1}_{kN} + \sum_{\alpha} m_\alpha \otimes W_\alpha \right]  = \int \dd \psi \dd \bar \psi \, e^{\bar \psi \psi + \sum_{\alpha} m_\alpha \bar \psi W_\alpha \psi} \, .
\end{equation}
Here the coefficients $m_\bullet$ are now $k \times k$ matrices and the matrix multiplication is implicit.

The mesons are now labeled by two extra indices i.e. $M_{\alpha;ij}$, $i,j=1,\dots,k$ representing $\bar\psi_i W_\alpha \psi_j$, while the $U(N)$ indices are summed over. Therefore each entry of $R$-generators now has two extra indices. We will find it more convenient to write the generalized Schwinger-Dyson equations in the dual presentation in Section \ref{subsec:multigg}.

In conclusion, we have found a neat ``generators and relations'' presentation $\cR_\leq$ of $\cH_{d;N}$ as a $\bC[T_\bullet]$-module, with generators $R_A$ with $|A|\leq N$ and relations (\ref{eq:role}). In particular, trace relations have been absorbed into the larger collection of relations satisfied by $T$'s and $R$'s. See Appendix \ref{app:SD} for examples. In the next Section, we will discuss a stringy interpretation of this presentation.\footnote{The idea of studying $\cH_{d;N}$ as a $\bC[T_\bullet]$-module is inspired by the work of \cite{Lee:2023iil}, which explored possible resolutions of $\cH_{d;N}$ as a complex of free modules.} 

\subsection{Fermionic Matrices}
\label{subsec:fermions}

This analysis can be extended to the space $\cH_{d,f;N}$ of $U(N)$-invariants built from $d$ bosonic and $f$ fermionic matrices.

As an example, we consider $U(N)$-invariant polynomials in a single fermionic matrix $\Psi$, reviewed in detail in \cite{Eleftheriou:2026yct}. Graded cyclicity of the trace implies that $\Tr\Psi^{2k}=0$, for $k\geq 1$, while the odd traces anticommute and square to zero. At large $N$, there are no further relations. We can thus introduce variables $T_{2k+1}$ and write
\begin{equation}
    \cH_{0,1;\infty} = \bC[T_1,T_3,\dots] \, .
\end{equation}

At finite $N$, the matrix satisfies $\Psi^{2N}=0$, so the odd traces $\Tr \Psi^{2N+2k+1}$ for $k\geq 0$ vanish. It is well known that these are the only trace relations. We therefore have
\begin{align}
    \mathcal{H}_{0,1;N} = \mathbb{C}[T_1,T_3,\dots,T_{2N-1}] \, .
\end{align}

We can also define words $W_k = \Psi^k$, and define $\Lambda_A$ by the usual auxiliary Grassmann integral, using mesons $\bar \psi W_k \psi$. The order of $\alpha_i$ in $A=(\alpha_1, \cdots, \alpha_n)$ only matters up to a sign, as $\Psi^k$ is fermionic for odd $k$. So e.g. $\Lambda_{(1,3)} = - \Lambda_{(3,1)}$. We also get a factor $\Lambda_{(\alpha)}=(-1)^{|\alpha|}T_{[\alpha]}$. Some of the $\Lambda_A$ will vanish identically for all $N$, e.g. $\Lambda_{(2k)}=\Tr \Psi^{2k}=0$. It turns out this condition can be derived from the open loop equations, so we do not need to impose it by hand.

The open loop equations also involve appropriate Koszul signs, which are instrumental in setting to $0$ certain $\Lambda_A$. Concretely, this results into an extra sign $(-1)^{|\alpha|}$ for the $T \Lambda$ term, and a $(-1)^{|\alpha||\beta_j|}$ sign for all the $\beta_j$ to the left of $\alpha$, including $\beta_i$ for the second open loop equation. For example,
\begin{align}
\begin{split}
    -T_1 \Lambda_{(2k-1)} &=\Lambda_{(1,2k-1)} + \Lambda_{(2k)} \\
    -T_1 \Lambda_{(2k-1)} &=\Lambda_{(1,2k-1)} - \Lambda_{(2k)} \, ,
\end{split}
\end{align}
which in particular implies $\Lambda_{(2k)}=0$. 

We can then define as before $\bC[T_\bullet]$ modules $\cR$, $\cR_\leq$ and $\cR_>$
and ask analogous questions as in the bosonic case:
\begin{itemize}
    \item Do the OLE reconstruct the $\Lambda_A$ as polynomials in $T_\bullet$? Equivalently, is $\cR \simeq \bC[T_\bullet]$?
    \item Is $\cR_\leq$ isomorphic to the space $\cH_{0,1;N}$ of invariants built from $\Psi$?
    \item Is $\cR_>$ isomorphic to the space of trace relations $\mathcal{T}_{0,1;N}$?
\end{itemize}
It seems clear that $\cR \simeq \bC[T_\bullet]$, as the OLE manifestly recast a $\Lambda_A$
with $(n+1)$ entries as a function of $T_\bullet$ and shorter $\Lambda$'s. Our argument for 
$\cR_\leq \simeq \cH_N$ also seems to go through as in the bosonic case. What we can only test numerically is the relationship between $\cR_>$ and the space of trace relationships.

\section{Giant Graviton Moduli Space}
\label{sec:ggmoduli}

In the context of the 't Hooft expansion, determinant operators 
become D-branes in a dual string theory description. In holographic examples with a known string theory dual, these are ``Giant Graviton'' D-branes, i.e. D-branes which reach the holographic boundary at a point. There is a rich literature \cite{Biswas:2006tj,Ashok:2008fa,Pasukonis:2012zj,Beasley:2002xv} which aims to refine this statement to a match between states obtained by a semi-classical quantization of Giant Graviton D-branes and analogues of the $\Lambda_A$ operators. 

Accordingly, we propose a powerful interpretation of the $\cH_{d;N} \simeq \cR_{\leq }$ isomorphism: gauge theory states at finite $N$ can be fully accounted for holographically by a judiciously gauged quantization of Giant Gravitons dressed by closed strings. This includes operators built from large numbers of matrices, even scaling as $N^2$ or worse. 
In this formulation, the OLE become the constraints associated to the worldvolume gauge symmetry of the Giant Graviton D-branes.

Previous work on the holographic interpretation of Giant Gravitons observed that the background of the original stack of $N$ D-branes induces an effective magnetic field on the moduli space of the Giant Gravitons \cite{Beasley:2002xv,Biswas:2006tj}. It was also observed that BPS/protected states would arise from the lowest Landau level for this magnetic field, potentially leading to a geometric quantization of the Giant Graviton moduli space \cite{Beasley:2002xv,Biswas:2006tj,Mandal:2006tk,Martelli:2006vh,Bhattacharyya:2007sa,Ashok:2008fa,Pasukonis:2012zj,Bergman:2012qh,Eleftheriou:2023jxr,Lee:2024hef}.

A classic example is the moduli space of $1/2$-BPS Giant Gravitons in ${\cal N}=4$ SYM: the support of the D-branes can be described as the intersection of a complex hypersurface $x=x_0$ in $\bC^3$ and the unit $S^5$ defined by $|x|^2 + |y|^2 + |z|^2 =1$ \cite{Mikhailov:2000ya}. The ``boundary'' at $|x_0|=1$ represents infinitesimally small D-branes and is naturally shrunk to a point, giving a moduli space with the geometry of $\bC P^1$. Geometric quantization of $\bC P^1$ with a symplectic form of order $N^{-1}$ reproduces states to be identified with subdeterminant operators, the coefficients of $\det(m-X)$ \cite{Biswas:2006tj}.

This construction was tentatively extended to $1/8$-BPS states using more general polynomial hypersurfaces $m(x,y,z)=0$ \cite{Mikhailov:2000ya}, but the extension raised important concerns \cite{Biswas:2006tj}. In particular, the moduli space naturally includes factorized loci with equation $m_1(x,y,z)m_2(x,y,z)=0$,
leading to a potential overcounting if one attempts to ``second-quantize'' the contribution of ``individual'' Giant Gravitons. Furthermore, the corners of moduli space corresponding to ``small'' D-branes reproduce states which are usually associated to closed strings instead. 

We propose that these issues can be understood as gauge artifacts: the moduli space of Giant Graviton D-branes is best described as a gauge quotient of a larger space including closed strings, and the gauge symmetries can convert open string excitations into closed strings. Our proposal will be mostly motivated from the gauge theory side of the duality, but in later Sections we will see the dual D-branes emerge at least at tree level in twisted holography.  

\subsection{Generating Functions}
In this section, we will often work with the dual space $\cH_{d;N}^\vee$ which is more readily identified as the result of a geometric quantization. 

First, consider the generating function of polynomials in single-trace operators
\begin{align}
    \cZ_0(t_\bullet) \equiv \exp \sum_{[\alpha]} t_{[\alpha]} \Tr W_\alpha \, .
\end{align}
Because of the trace relations, $\cZ_0(t_\bullet)$ is not a generic function of the $t_\bullet$ variables. Instead, it defines a paring 
between $\cH_{d;N}$ and a dual linear space of polynomials which can be identified with $\cH_{d;N}^\vee$. For example, all traces vanish for $N=0$ and $\cZ_0(t_\bullet)$ is $t_\bullet$-independent: $\cH_{d;0}^\vee$ is generated by the polynomial ``$1$''.

Analogously, the generating function 
\begin{equation}
    \cZ_1(t_\bullet,m_\bullet) \equiv \exp\qty( \sum_{[\alpha]} t_{[\alpha]} \Tr W_\alpha ) \det_{N \times N}\!\left[m_0 \mathbf{1}_N + \sum_\alpha m_\alpha W_\alpha \right]
\end{equation}
presents $\cH_{d;N}^\vee$ as a subspace of the space of polynomials in $t_\bullet$ and $m_\bullet$, of homogeneous degree $N$ in $m_\bullet$. From now on, we include $m_0$ in the notation $m_\bullet$ and write $m_{\bullet \neq 0}$ if we need to explicitly exclude it.

The open loop equations can be presented as differential equations satisfied by $\cZ_1(t_\bullet,m_\bullet)$: 
\begin{equation}
\begin{split}
    \cD_{+,\alpha} \cZ_1(t_\bullet,m_\bullet) &=\left(\partial_{t_{[\alpha]}} - m_0 \partial_{m_\alpha}- \sum_\beta m_\beta \partial_{m_{\alpha \beta}}\right)\cZ_1(t_\bullet,m_\bullet) =0\cr
    \cD_{-,\alpha} \cZ_1(t_\bullet,m_\bullet) &=\left(\partial_{t_{[\alpha]}} - m_0  \partial_{m_\alpha} - \sum_\beta m_\beta \partial_{m_{\beta\alpha }} \right)\cZ_1(t_\bullet,m_\bullet)=0 \, .
\end{split}
\end{equation}
The isomorphism $\cH_{d;N} \simeq \cR_\leq$ can be stated dually as the isomorphism between $\mathcal{H}_{d;N}^\vee$ and the space of polynomial solutions of the open loop differential constraints, of homogeneous degree $N$ in $m_\bullet$:
\begin{equation}
    \ker_{\cD_{\pm,\bullet}} \bC[t_\bullet,m_\bullet] = \bigoplus_{N=0}^\infty \mathcal{H}_{d;N}^\vee \, ,
\end{equation}
where $N$ is the total degree in the $m_\alpha$ variables. Equivalently, we can write
\begin{framed}\begin{equation}
    \cH_{d;N}^\vee \simeq \cR_\leq^\vee = \ker_{\cD_{\pm,\bullet}} \bC[t_\bullet,m_\bullet]^N \, .
\end{equation}\end{framed}

We are ready to give a stringy interpretation to this presentation. The $m_\bullet$ variables modulo overall scale parameterize a $\bC P^\infty$ moduli space of determinant operators and thus formally of Giant Graviton D-branes. Homogeneous polynomials of degree $N$ in the $m_\bullet$ are precisely the result of a geometric quantization of such a moduli space. The factor of $\bC[t_\bullet]$ dresses the outcome of the geometric quantization by a Fock space of closed string states. Our next goal will be to interpret the open loop equations as {\it gauge constraints} for the open string degrees of freedom on the Giant Graviton brane. 

We can give a description of the trace relations in the same language. We can present $\cR^\vee$ as
\begin{equation}
    \cR^\vee = \ker_{\cD_{\pm,\bullet}} \bC[t_\bullet,m_\bullet,m_0^{-1}]^N \, .
\end{equation}
Indeed, allowing negative powers of $m_0$ also allows the degree in the remaining $m_\alpha$ to be greater than $N$. 

The well known ``residue map'' exact sequence 
\begin{equation}
    0\to \bC[z] \to \bC[z,z^{-1}]\to \bC[\partial_z] \delta(z) \to 0 \, ,
\end{equation}
then allows us to tentatively conjecture 
\begin{mdframed}[
    linecolor=none,
    tikzsetting={draw=black,dashed,line width=0.4pt},
    innertopmargin=12pt,
    innerbottommargin=15pt
]
\begin{equation}
    {\cal T}_{d;N}^\vee \simeq \mathcal{R}_>^{\vee} = \ker_{\cD_{\pm,\bullet}} \left[\bC[t_\bullet,m_{\bullet\neq 0} ,\partial_{m_0}]\delta(m_0) \right]^N \, .
\end{equation}
\end{mdframed}
Geometrically, the large $N$ answer $\cR^\vee$ appears as a geometric quantization of the $m_0 \neq 0$ locus in the $\bC P^\infty$ moduli space and the finite $N$ answer corresponds to wave-functions which are regular at $m_0$.

\subsection{Multiple Giant Gravitons}
\label{subsec:multigg}

In a string theory setting, it would be natural to include any number of Giant Graviton D-branes. We therefore consider a generalization to ``$k$-determinant'' operators built from $k$ pairs of $\epsilon$ tensors:
\begin{equation}
    \cZ_k(t_\bullet,m_\bullet) \equiv \cZ_0(t_\bullet)  \det_{kN \times kN} \left[m_0 \otimes \mathbf{1}_N + \sum_\alpha m_\alpha \otimes W_\alpha\right] \, .
\end{equation}
Here $m_0,m_\alpha$ are $k\times k$ matrices, and the tensor product yields $kN\times kN$ matrices.
The expansion of such an expression involves $U(k) \times U(k)$-invariant polynomials of degree $k N$ in the $m_\bullet$. 

This generating functions satisfies matrix-valued open loop equations
\begin{align}
    \cD_{+,\alpha} \cZ_k(t_\bullet,m_\bullet) &=\left(\partial_{t_{[\alpha]}} - m_0 \partial_{m_\alpha}- \sum_\beta m_\beta \partial_{m_{\alpha \beta}}\right)\cZ_k(t_\bullet,m_\bullet) =0\cr
    \cD_{-,\alpha} \cZ_k(t_\bullet,m_\bullet) &=\left(\partial_{t_{[\alpha]}} -  \partial_{m_\alpha}m_0  - \sum_\beta \partial_{m_{\beta\alpha }} m_\beta \right)\cZ_k(t_\bullet,m_\bullet)=0 \, ,
\end{align}
where matrix multiplication is now implied. 

The Grassmann integral satisfies two extra constraints, which we did not include among the $\cD_{\pm,\alpha}$ as our notation excludes $\alpha = \emptyset$: the transformation $m_\bullet \to g_\ell m_\bullet g_r$ multiplies $\cZ_k(t_\bullet,m_\bullet)$ by $(\det^N g_\ell)(\det^N g_r)$. Infinitesimally, 
\begin{align}
    \cD_{+,0} \cZ_k(t_\bullet,m_\bullet) &=\left(N - m_0 \partial_{m_0}- \sum_\beta m_\beta \partial_{m_{ \beta}}\right)\cZ_k(t_\bullet,m_\bullet) =0\cr
    \cD_{-,0} \cZ_k(t_\bullet,m_\bullet) &=\left(N -  \partial_{m_0}m_0  - \sum_\beta \partial_{m_{\beta}} m_\beta \right)\cZ_k(t_\bullet,m_\bullet)=0 \, ,
\end{align}
require $\cZ_k(t_\bullet,m_\bullet)$ to transform in the $N$-th power of the determinant representation of $U(k) \times U(k)$.

We can thus represent the dual space $\mathcal{H}^\vee_{d;N}$ as
\begin{framed}\begin{equation}
    \cH_{d;N}^\vee \simeq \ker_{\cD_{\pm,\bullet}} \bC[t_\bullet,m_\bullet]^{U(k)_N \times U(k)_N} \, ,
\end{equation}\end{framed}
\noindent for all $k$ with $m_\bullet$ being $k \times k$ matrices and $U(k)_N$ denoting a projection on the $\det^N$ irrep. The proof is a simple generalization of the $k=1$ case. 

This identification actually becomes more transparent as we increase $k$. If we focus on invariant polynomials in the $X_i$ of degree up to $k$, the leading part of the $k$-determinant 
\begin{equation}
    \det_{kN \times kN} \left[m_0 \otimes \mathbf{1}_N + \sum_i m_i X_i\right] \, ,
\end{equation}
already gives a nondegenerate pairing between the invariant polynomials and appropriate expressions built from $m_\alpha$ with $|\alpha| \leq 1$. Essentially, modifications of $(\det m_0)^N$ obtained by removing a finite number of $m_0$'s and attaching $m_i$ in various ways to the dangling indices of the $N$ determinants. 

As the determinant can be expanded into traces by inverting $m_0$, and $m_0$ can be inverted as long as $\det m_0$ can be, the analogue to $\cR^\vee$ is 
\begin{equation}
    \ker_{\cD_{\pm,\bullet}} \bC[t_\bullet,m_\bullet,\det m_0^{-1}]^{U(k)_N \times U(k)_N} 
\end{equation}
and we expect it to be isomorphic to $\bC[t_\bullet]$. There is a ``residue map'' exact sequence 
\begin{equation}
    \bC[t_\bullet,m_\bullet] \to \bC[t_\bullet,m_\bullet,\det m_0^{-1}]\to \bC[t_\bullet,m_\bullet,\partial]\delta(\det m_0) \, .
\end{equation}
Here $\partial$ denotes derivatives of the delta function. This conjecturally gives us a candidate for the trace relations
\begin{mdframed}[
    linecolor=none,
    tikzsetting={draw=black,dashed,line width=0.4pt},
    innertopmargin=12pt,
    innerbottommargin=15pt
]\begin{equation}
    \mathcal{T}_{d;N}^\vee \simeq \ker_{\cD_{\pm,\bullet}} \qty[\bC[t_\bullet,m_\bullet,\partial]\delta(\det m_0)]^{U(k)_N \times U(k)_N} \, .
\end{equation}
\end{mdframed}

\subsection{The Algebra of Open Loop Equations}
\label{subsec:OLEalgebra}

The interpretation of the open loop equations as gauge constraints is reinforced by the observation that the associated differential operators $\cD_{\pm,\bullet}$ form a Lie algebra. For $k=1$ we compute
\begin{align}
    [\cD_{+,\alpha}, \cD_{+,\beta}] &= -\cD_{+,\beta \alpha}+ \cD_{+,\alpha \beta} \nonumber \\
    [\cD_{+,\alpha}, \cD_{-,\beta}] &= [\cD_{-,\alpha}, \cD_{+,\beta}] =0 \\
    [\cD_{-,\alpha}, \cD_{-,\beta}] &= -\cD_{-,\alpha \beta} + \cD_{-,\beta \alpha} \, . \nonumber
\end{align}
This gives two copies of the Lie algebra associated to the tensor algebra
\begin{align}
    T^\bullet \mathbb{C}^d = \bigoplus_{n=0}^\infty (\mathbb{C}^d)^{\otimes n} \, ,
\end{align}
consisting of linear combinations of the words $W_\alpha$ in $d$ matrices, with product  $W_\alpha W_\beta=W_{\alpha\beta}$. For general $k$ we get 
\begin{equation}
    \fgl_k[T^\bullet \bC^d] \oplus \fgl_k[T^\bullet \bC^d] \, ,
\end{equation}
where $\fgl_k[\cA]$ is the Lie algebra of $k \times k$ matrices valued in an algebra $\cA$.

The open loop equations also suffer a certain degree of redundancy. In particular,
\begin{equation} \label{eq:gaugegauge}
    \sum_\alpha \cD_{+,\alpha \beta} m_\alpha + \cD_{+,\beta} m_0  = \sum_\gamma m_\gamma \cD_{-,\beta \gamma} + m_0 \cD_{-,\beta} \, .
\end{equation}
The $\cD m$ term is a short hand for $(m^T \cD^T)^T$, i.e. the derivative acts before multiplication by $m$, but the matrix multiplication happens in the opposite order. 

We can think about this relation as a gauge symmetry for the gauge generators themselves. A more familiar example of this phenomenon is the case of a 1-form symmetry, where the gauge generator is itself a connection. This also happens naturally in string field theory. 

\subsection{The Gauge Symmetry of Complexes}
We should discuss an instructive toy model with a similar gauge structure. Consider the simplest type of a complex: a pair of vector spaces $V_0$ and $V_1$ assigned ghost numbers $0$ and $1$ respectively, together with a map $d: V_0 \to V_1$ playing the role of a differential. 

  would we define a moduli space of such differentials? It would certainly be natural to identify differentials related by a base change in $V_0$ and $V_1$. Infinitesimally, this gives a ``gauge symmetry''
\begin{equation}
    d \to \epsilon_\ell d -d \epsilon_r \, .
\end{equation}
Some of these gauge transformations, though, are automatically trivial: we can pick any $\rho:V_1\rightarrow V_0$ and shift 
\begin{equation}
    \epsilon_\ell \to d \rho \qquad \qquad \epsilon_r \to \rho d
\end{equation}
without changing the variation of $d$. It is natural to treat this as a gauge symmetry for the gauge generators themselves. 

The formal similarity of this problem to the open loop equations should be obvious: the $\cD_{\pm,0}$ differential operators 
generate transformations 
\begin{equation}
    m_0 \to \epsilon_\ell m_0 -m_0 \epsilon_r
\end{equation}
of $m_0$. 

The similarity grows if we consider position-dependent complexes such as these employed to describe B-model D-branes e.g. in $\bC$.
There $d$ is a matrix-valued polynomial and so are the gauge parameters. If we expand 
\begin{equation}
    d = \sum_n d_n z^n \, ,
\end{equation}
the individual coefficients transform as 
\begin{equation}
    d_n \to \sum_{n'} \epsilon_{\ell,n'} d_{n-n'} -d_{n-n'} \epsilon_{r,n'} \, .
\end{equation}
This is analogous to the action of $\cD_{\pm,\alpha}$ on $m_\beta$, 
with the difference that the composition of words $\alpha \beta$ is replaced by the commutative product of monomials $z^{n+n'}$.

From that perspective, the open loop equations take the form of gauge symmetries for a complex whose differential
\begin{equation}
    d = m_0 + \sum_\alpha m_\alpha W_\alpha
\end{equation}
is valued in the noncommutative algebra $T^\bullet \bC^d$, perhaps thought of as the algebra of functions on some noncommutative version of a space. Turning on a tree-level differential in conventional supersymmetric gauge theory examples will give an actual space. In the geometric analogy, a complex with $\mathrm{dim} V_0 - \mathrm{dim} V_1 = 0$ represents a co-dimension $1$ D-brane located at $\det d=0$.

Our presentation of $\cH_{d;N}^\vee$ thus appears to morally describe a gauged geometric quantization of a moduli space of ``complexes''
\begin{equation}
    \bC^k \to_d \bC^k \, .
\end{equation}

Analogous considerations apply to $\cH_{d,f;N}^\vee$. We will momentarily include the effect of 
interactions at tree level, which will make the geometric interpretation of this construction more transparent. 

\section{BRST Complex for Open Loop Equations}
\label{sec:QOLE}

The goal of this Section is to study the effect of trace relations in situations where $\cH_N$ is defined by a cohomology problem, as for the case of BPS subsectors of supersymmetric gauge theories. In the process, we will find it natural to organize the open loop equations into a complex as well, with clear connections to open string field theory and to categorical descriptions of Giant Graviton D-branes.

\subsection{Lie Algebra Cohomology}

Many problems involving BPS operators define the analogue of $\mathcal{H}_N$ as a complex, with a supercharge acting as a differential $Q$ on $U(N)$-invariant polynomials in a collection of matrices of various ghost number and fermionic parity. We will denote these matrices collectively as $\Phi_i$.

The action of $Q$ always contains a tree-level term\footnote{Depending on the specific quantum field theory and BPS sector, there may also be loop corrections to $Q$ with a more intricate action \cite{Budzik:2023xbr, Choi:2025bhi, Budzik:2025zvu}. In this paper we will only consider tree-level differentials and leave loop corrections for future work.}, which maps fields to words in the fields:
\begin{equation} \label{eq:QDef}
    Q \Phi_i = \sum_\alpha q_i^\alpha W_\alpha \, ,
\end{equation}
and acts on polynomials in the matrix elements by the Leibniz rule.
 
We will now translate some of these definitions to the language of Homological Algebra, following string field theory literature \cite{Witten:1992yj,Zwiebach:1992ie,Lada:1992wc,Alexandrov:1995kv,Gaberdiel:1997ia,Kontsevich:1997vb,Kontsevich:2006jb,Hohm:2017pnh,Jurco:2018sby}. 

The first entry in the dictionary is the notion of an $\cA_\infty$ algebra. Albeit both the name and many traditional presentations of $\cA_\infty$ algebras point to a generalization of ``algebra'', an $\cA_\infty$ algebra $\cA$ is best defined as a differential mapping letters to words as in (\ref{eq:QDef}).\footnote{More precisely as a formal pointed noncommutative dg-manifold with coordinates $\Phi_i$.} Given an actual associative algebra $\cA$ we can always consider a field $\Phi$ of ghost number $1$ valued in $\cA$ with differential 
\begin{equation}
    Q \Phi = \Phi^2 \, .
\end{equation}
If we pick a basis $a_i$ in $\cA$ and expand 
\begin{equation}
    \Phi = \sum_i \Phi_i a^i \, ,
\end{equation}
the structure constants of $\cA$ become coefficients for a general differential quadratic in the $\Phi_i$, and viceversa. A dg-algebra can be similarly associated to a differential with a linear term:
\begin{equation}
    Q \Phi = d \Phi + \Phi^2 \, .
\end{equation}

In situations where we can identify the $\Phi_i$ as the fields of an open string field theory, the $a_i$ can be thought of as open vertex operators and $\cA$ as the $\cA_\infty$ algebra of open strings. In typical examples, the gauge theory of interest is the worldvolume theory of some D-branes, the BPS operators are local operators in a twisted version of the gauge theory and 
$\cA$ is the $\cA_\infty$ algebra of open strings in the analogous twist of the string theory. 

Before getting to concrete examples, it is useful to mention a small subtlety concerning the interplay between $U(N)$ invariance and cohomological constructions. The $\cA_\infty$ algebra $\cA$ in examples will be usually equipped with an unit and an augmentation, so that we can expand 
\begin{equation}
    \Phi = C + \sum_i \Phi_i a^i \, ,
\end{equation}
with a ``ghost zeromode'' $C$ and a differential of the form 
\begin{equation}
    Q \Phi_i = [C,\Phi_i] + Q_r \Phi_i \, , \qquad \qquad Q C = \frac12 [C,C]\, .
\end{equation}
Here $[\bullet,\bullet]$ denotes graded commutators and $Q_r \Phi_i$ does not involve $C$. 

The complex of polynomials in the matrix elements of $C$ and $\Phi_i$ with differential $Q$ is the absolute Chevalley-Eilenberg complex $C^\bullet(\mathfrak{gl}_N[\cA])$. Its cohomology is called absolute cohomology and is denoted by $H^\bullet(\mathfrak{gl}_N[\cA])$.
The relative Chevalley–Eilenberg complex $C^\bullet(\fgl_N[\cA]|\fgl_N)$
denotes the complex of $U(N)$-invariant polynomials of $\Phi_i$ with differential $Q_r$,
which is what we are typically interested in:  
\begin{equation}
    \cH_{\cA;N} \equiv H^\bullet(\fgl_N[\cA]|\fgl_N) \, .
\end{equation}
The absolute and relative cohomologies are related by
\begin{align}
H^\bullet(\mathfrak{gl}_N[A])
\simeq \mathcal H_{A;N}
\otimes \mathbb{C}[\Tr C,\Tr C^3,\dots,\Tr C^{2N-1}] \, .
\end{align}

In our computations, we will mostly use three examples. In each case, the differential is determined by $Q\Phi=\Phi^2$:
\begin{itemize}
    \item A single bosonic matrix\footnote{The original $d$-matrix problem is encoded by the algebra $\cA=\mathbb{C}\oplus\mathbb{C}^d[1]$, which is equivalent to $\mathbb{C}[\theta]$ for $d=1$.}  $X$ is described by $\cA=\bC[\theta]$, where $\theta$ is fermionic and has ghost number 1. The superfield is
    \begin{align}
    \Phi &= C + X \theta \, .
    \end{align}

    \item The $1/4$-BPS subsector of $\mathcal N=4$ SYM is described by $\mathcal A=\mathbb C[\theta_1,\theta_2]$, where both $\theta_i$ are odd of ghost number 1. The superfield contains bosonic matrices $X,Y$ and a fermionic matrix $B$:
    \begin{align}
        \Phi=C+\theta^1 X + \theta^2 Y + \theta^1\theta^2 B \, .
    \end{align}
    In particular,
    \begin{align}
    QC &=C^2 \\
    QX &=[C,X] \, , \qquad QY=[C,Y] \\
    Q\Psi &=[C,\Psi]+[X,Y] \, .
    \end{align}

    \item A fermionic matrix $\Psi$ of ghost number $1$ and its derivatives are described by $\mathcal A=\mathbb C[z]$, where $z$ is even of ghost number $0$. We write
    \begin{equation}
    \Phi = C(z) = C + \sum_{n=1}^\infty \frac{z^n}{n!} \partial^n C
    \end{equation}
    and $QC(z)=C(z)C(z)$. This reproduces the ``dual $1/2$-BPS sector'' of $\mathcal{N}=4$ SYM, with $\partial C$ being a gaugino. 
\end{itemize}
Other examples of note involve the BMN sector of ${\cal N}=4$ SYM asssociated to $\cA=\bC[\theta_1,\theta_2, \theta_3]$ as well as the full $1/16$-BPS cohomology $\cA=\bC[\theta_1,\theta_2, \theta_3,z_1,z_2]$. All of these algebras occur on D-branes supported on some $\bC^{d_2}$ 
in a B-model topological string theory with target space $\bC^{d_1+d_2}$. 

The first payoff of the Homological Algebra language are theorems which constrain the structure of the large-$N$ limit. They should be understood as refined combinatorial results with a transparent interpretation in holography and string theory. 

The action of $Q$ is clearly defined in a manner which does not depend on the $\Phi_i$ being matrices of some fixed size $N$: it is uniform in $N$. Accordingly, we can define a linear action of $Q$ on abstract words $W_\alpha$ and on the abstract $T_\bullet$ representing single-traces $\Tr W_\alpha$. The action of $Q$ on $T_\bullet$ defines the so-called {\it relative cyclic cohomology} complex $CC_r^{\bullet-1}(\cA)$
with cohomology denoted as $HC_r^{\bullet-1}(\cA)$. We can thus identify the cohomology of $(\bC[T_\bullet],Q)$ as
\begin{equation}
    \cH_{\cA;\infty} = S^\bullet \left(HC_r^{\bullet-1}(\cA) \right)\, . 
\end{equation}
This is referred to as the LQT theorem \cite{Tsygan:1983,Loday:1984} in the math literature.

There are useful mathematical tools to compute $HC_r^{\bullet-1}(\cA)$ for algebras of interest, and the relation between $Q$-cohomology in supersymmetric gauge theories and twisted holography \cite{Costello:2018zrm, Budzik:2023xbr, Gaiotto:2024dwr} often reflects in natural identifications between  $HC_r^{\bullet-1}(\cA)$ and spaces of closed strings in an holographic dual description of the theory.

\subsection{Words and Koszul Duality}

The action of $Q$ on the space of words is also very interesting, and relevant for the study of mesons. The action of $Q$ is compatible with concatenation of words, so words form a dg-algebra. If we include the trivial word as the identity, this dg-algebra is known as the Koszul dual $\cA^!$ to $\cA$. If we exclude the identity, we get the augmentation ideal $\mathcal{A}^!/\bC$. A striking result is that this is literally a duality up to quasi-isomorphism: $(\cA^!)^!\simeq \cA$. 
Furthermore, it behaves well under tensor product:
\begin{equation}
    (\cA \otimes \mathcal{B})^! \simeq \cA^! \otimes \mathcal{B}^! \, .
\end{equation}
In string field theory language, if $\cA$ is the algebra of open string vertex operators for some D-brane $\cD$, then $\cA^!$ describes open string modes for a D-brane $\cD^!$ which is ``fully transverse'' to $\cD$, i.e. such that there is a unique boundary-changing local operator between the $\cD$ and $\cD^!$ and viceversa. Intuitively, the description encodes the open string modes of $\cD^!$ via mixed disk amplitudes with $\cD$ open string insertions.

In all the examples above, $\cA^!$ is quasi-isomorphic to a much simpler algebra. In particular, for $\cA=\mathbb{C}[\theta]$ the Koszul dual algebra is $\cA^!\simeq\mathbb{C}[x]$ for bosonic $x$. 
These correspond to Dirichlet and Neumann D-branes in a B-model with target $\bC$. Taking tensor products, if $\cA = \bC[\theta^1,\cdots, \theta^{d_1}, z^1, \cdots z^{d_2}]$ then $\cA^! \simeq \bC[x_1,\cdots, x_{d_1}, \zeta_1, \cdots \zeta_{d_2}]$, with conventional bosonic variables $x_i$ and fermionic variables $\zeta_j$ of ghost number $1$. These are B-model D-branes wrapping $\mathbb{C}^{d_2}$ in $\mathbb{C}^{d1+d2}$ and  $\mathbb{C}^{d_1}$ in $\mathbb{C}^{d_1+d_2}$ respectively.

\subsection{The BV Complex of Mesons}
\label{sec:complex}

Given some polynomial $P(\Tr W_\bullet,\bar{\psi} W_\bullet\psi)$,
transforming in the $\det^N \times \det^N$ representation of $U(k) \times U(k)$, 
the Grassmann integral 
\begin{equation}
    \int \dd \psi \dd \bar \psi\, P(\Tr W_\bullet,\bar \psi W_\bullet \psi) 
\end{equation}
produces an $U(N)$-invariant polynomial in the fields $\Phi_i$. We can bring the action of $Q$ through the integral
\begin{equation}
    \int \dd \psi \dd \bar \psi \,Q P(\Tr W_\bullet,\bar \psi W_\bullet \psi) 
\end{equation}
and the image will be in $Q$-cohomology if $Q P(\Tr W_\bullet,\bar \psi W_\bullet \psi)$
is a total derivative. Our discussion in the previous sections indicates that 
at this stage we do not need to worry about trace relations anymore, in the sense that 
elements which vanish by trace relations can be rewritten as total derivatives as well. 

We will now show how the condition of being a total derivative can be expressed in terms of the abstract $T_\bullet$ and $M_\bullet$ variables as well, and how to encode $N$ fully in 
the restriction to polynomials in the $\det^N \times \det^N$ representation. 

We will make use of the BV complex for the Grassmann integral \eqref{eq:grass}, which imposes the Schwinger-Dyson equations in a cohomological way. Concretely, one introduces \textit{anti-fields} $u$ and $\bar u$ for each of the $\psi$ and $\bar \psi$ variables, with the same quantum numbers but ghost number $-1$, and a nilpotent differential
\begin{equation}
    \Delta_\BV = \Tr_N \partial_{\bar \psi} \partial_u  + \Tr_N \partial_{\bar u} \partial_\psi \, ,
\end{equation}
called the \textit{BV Laplacian}, which raises the ghost number by 1. This imposes in a derived way the vanishing of total derivatives, i.e.
\begin{equation}
    \Delta_\BV \left(f u\right) =\partial_{\bar \psi} f \, , \qquad \qquad \Delta_\BV \left(\bar u f\right) = \partial_{\psi} f \, . \label{eq:derivatives}
\end{equation}

The failure of $\Delta_{\BV}$ to act as a derivation defines the BV bracket
\begin{equation}
    \{A,B\}_\BV \equiv \Delta_\BV (AB) - (\Delta_\BV A)B - (-1)^{|A|} A(\Delta_\BV B) \, , \label{eq:bracket}
\end{equation}
i.e.
\begin{align}
\begin{split}
    \{A,B\}_\BV &= \Tr_N \partial_{\bar \psi} A \partial_u B+(-1)^{|A||B|} \Tr_N \partial_{\bar \psi}B \partial_u A + \cr &+(-1)^{|A||B|+|B|} \Tr_N \partial_{\bar u} B \partial_\psi A+ (-1)^{|A|}\Tr_N \partial_{\bar u} A\partial_\psi B \, .
\end{split}
\end{align}
The BV bracket behaves as a shifted version of a Poisson bracket: it maps functions to vector-fields with ghost number increased by one as $A \to \{A,\bullet\}_\BV$. 
The BV differential is a derivation of the bracket as well.

Integration is defined by picking a polarization, i.e. setting $u=\bar u = 0$, and integrating over $\psi$, $\bar \psi$ as usual. Then obviously $\Delta_{\BV}$-exact elements integrate to zero:
\begin{equation}
    \int \dd \psi \dd \bar \psi \, \Delta_{\BV} (\cdots) =0 \, ,
\end{equation}
as this can only be non-vanishing when the ellipsis is linear in $u$ or $\bar u$, and then the argument of the integral is a total derivative due to \eqref{eq:derivatives}. Alternatively, one can say that the cohomology of $\Delta_\BV$ is one-dimensional, with the only class being the product of all fermionic variables, and the Grassmann integral picks out the coefficient of that class.

All of these calculations employ the extended set of mesons, to which we can associate formal expressions $\cM_\bullet = (M_\bullet, M^\pm_\bullet, M^r_\bullet)$: 
\begin{equation}
    M_\alpha \to \bar \psi W_\alpha \psi \qquad 
    M^+_\alpha \to \bar \psi W_\alpha u \qquad
    M^-_\alpha \to \bar u W_\alpha \psi \qquad
    M^r_\alpha \to \bar u W_\alpha u \, ,
\end{equation}
which represent them at infinite $N$. The BV bracket closes on the mesons in an uniform-in-$N$ manner. We can therefore define analogously the action of $\Delta_\BV$ on the whole space of polynomials $\bC[T_\bullet, \cM_\bullet]$ of formal traces and mesons, with $N$ appearing only in the 
variation of $\cM_0$ zeromodes. Concretely, the action on the mesons and its lift to the formal variables are
\begin{equation}
\begin{aligned}
\Delta_{\mathrm{BV}}(\bar\psi W_\alpha u)
&=\Tr W_\alpha \, ,
& \Delta_{\mathrm{BV}} M_\alpha^+ &=T_{[\alpha]} \\
\Delta_{\mathrm{BV}}(\bar uW_\alpha\psi)
&=(-1)^{|\alpha|}\Tr W_\alpha \, ,
& \Delta_{\mathrm{BV}} M_\alpha^-
&=(-1)^{|\alpha|}T_{[\alpha]} \, .
\end{aligned}
\end{equation}
For the empty word and zero-modes we have
\begin{equation}
\begin{aligned}
\Delta_{\mathrm{BV}}(\bar\psi u)
&=N \, ,
&
\Delta_{\mathrm{BV}}M_0^+ &=N \\
\Delta_{\mathrm{BV}}(\bar u\psi)
&=N \, , & \Delta_{\mathrm{BV}}M_0^- &=N \, .
\end{aligned}
\end{equation}

Going back to the goal of describing $\cH_{\cA;N}$ in terms of polynomials $P$ such that $Q P(T_\bullet,M_\bullet)$ is a total derivative, we can now write that condition as 
\begin{equation}
    Q P(T_\bullet,M_\bullet) + \Delta_\BV P^{(1)}(T_\bullet,\cM_\bullet) =0
\end{equation}
with the right hand side linear in $u$ and $\bar u$, modulo 
exact expressions or total derivatives:
\begin{equation}
   P(T_\bullet,M_\bullet) = Q R(T_\bullet,M_\bullet) + \Delta_\BV R^{(1)}(T_\bullet,\cM_\bullet) \, ,
\end{equation}
which works because $(Q + \Delta_\BV)^2=0$.

As the cohomology of $\Delta_\BV$ acting on $\psi$, $\bar \psi$, $u$, $\bar u$ is one-dimensional and proportional to the product of al the fermions, we may hope that the $\Delta_\BV$ cohomology 
of $\bC[T_\bullet, \cM_\bullet]$ will also be supported in $\bC[T_\bullet,M_\bullet]$. If that was the case, we could then write $\Delta_\BV Q P^{(1)}(T_\bullet,\cM_\bullet) =0$ and derive 
\begin{equation}
    Q P^{(1)}(T_\bullet, \cM_\bullet) + \Delta_\BV P^{(2)}(T_\bullet,\cM_\bullet) =0
\end{equation}
etcetera, and for exact classes we would write 
\begin{equation}
   \Delta_\BV P^{(1)}(T_\bullet,\cM_\bullet) + Q\Delta_\BV R^{(1)}(T_\bullet,\cM_\bullet) =0
\end{equation}
implying 
\begin{equation}
    P^{(1)}(T_\bullet,\cM_\bullet) = Q R^{(1)}(T_\bullet,\cM_\bullet) +\Delta_\BV R^{(2)}(T_\bullet,\cM_\bullet)
\end{equation}
etc. This would identify $\cH_{\cA;N}$ with the cohomology of $Q + \Delta_\BV$ acting on 
\begin{equation}
    \bC[T_\bullet,\cM_\bullet] \, ,
\end{equation}
and we would have archived a description of $\cH_{\cA;N}$ where $N$ is just a numerical parameter. 

Explicit inspection, though, reveals that the cohomology of $\Delta_\BV$ is more intricate than one may naively expect. For example, we can compute the cohomology of $\bC[\cM_0]$ and compare it to the cohomology of $\bC[\psi, \bar \psi, u, \bar u]$. Although $(\det M_0)^N$ is certainly a cohomology class and maps to a multiple of the product of all fermions, we find an extra class in ghost number $1$ for $k=1$, extra classes of ghost numbers $1,3,4$ for $k=2$, etc. This tells us that ignoring the analogue of trace relations for the fermions and antifields and replacing them with mesons is not an harmless step. Getting rid of these ``spurious'' classes will be a recurring theme through the rest of the Section. 

\subsection{The Enlarged Symmetry Algebra}

We will now characterize the complex $(\bC[T_\bullet, \cM_\bullet], Q + \Delta_\BV)$. The action of the differential on a polynomial involves two types of contributions: $Q+ \Delta_\BV$ acting on a single variable or $\{\bullet, \bullet\}_\BV$ acting on a pair of variables due to \eqref{eq:bracket}. Furthermore, the action of $Q+ \Delta_\BV$ is linear on the variables up to the correction linear in $N$ for the $M^\pm_0$ variables.

This is literally the definition of the {\it Chevalley-Eilenberg homology complex} 
\begin{equation}
    C_\bullet(\cL,\Ber^N) \equiv (\bC[T_\bullet, \cM_\bullet], Q + \Delta_\BV)
\end{equation} 
for the dg-Lie algebra $\cL$ defined by the differential $Q + \Delta_\BV$ and bracket $\{\bullet, \bullet\}_\BV$ acting on the variables, after shifting the ghost number of the variables by $1$. In other words, $\cL$ has generators $sT_\bullet$ and $s\cM_\bullet$ whose ghost number is increased by $1$ compared to the original variables. The ``$\Ber^N$'' argument will be motivated later on, but for now just denotes the $M^\pm_0 \to N$ correction to the differential. 

We are left with the task of identifying $\cL$. We should first consider the meson zeromodes. The $\bar \psi u$ and $\bar u \psi$ modes generate the obvious $\fgl_k \times \fgl_k$ symmetries: 
\begin{equation}
    \{\bar \psi u, u\}_\BV = u \qquad \qquad \{\bar \psi u,\bar \psi\}_\BV = - \bar \psi \, ,
\end{equation}
and 
\begin{equation}
     \{\bar u \psi,\psi\}_\BV = - \psi \qquad \qquad \{\bar u \psi,\bar u\}_\BV = \bar u  \, ,
\end{equation}
which simultaneously rotate a block of fields and their anti-fields. The fact that $\Delta_\BV$ 
maps them to $N$ encodes the fermion number anomalies which require integrands to transform as $\det^N \times \det^N$. 

The remaining zeromodes $\bar u u$ and $\bar \psi \psi$ extend the symmetry algebra to a
graded version of $\fgl_{k|k}$:
\begin{equation}
    \{\bar \psi \psi,u\}_\BV = \psi \qquad \qquad \{\bar \psi\psi,\bar u\}_\BV = - \bar \psi \, ,
\end{equation}
and 
\begin{equation}
    \{\bar u u,\psi\}_\BV =  u \qquad \qquad \{\bar u u,\bar \psi\}_\BV = \bar u \, .
\end{equation}
grouping together fields and anti-fields into (anti)fundamental representations. Note that $\fgl_{k|k}$ usually refers to infinitesimal linear transformations of a superspace $\bC^{k|k}$, while here we have a graded vector space instead, with $\bC^k$ summands in ghost numbers $0$ and $-1$. The off-diagonal generators of this graded Lie algebra have ghost number $1$ and $-1$ respectively. We will abuse notation and use $\fgl_{k|k}$ to describe this symmetry, for lack of concise alternatives.

The rest of the mesons enlarge the symmetry algebra further. As we defined the Open Loop Equations, we already implicitly saw how the $\bar \psi W_\alpha u$ and $\bar u W_\alpha \psi$ mesons generate rotations valued in words: 
\begin{equation}
    \{\bar \psi W_\alpha u, u\}_\BV = W_\alpha u \qquad \qquad \{\bar \psi W_\alpha u,\bar \psi\}_\BV = - (-1)^{|\alpha|}\bar \psi W_\alpha \, ,
\end{equation}
and 
\begin{equation}
     \{\bar u W_\alpha \psi,\psi\}_\BV = - (-1)^{|\alpha|} W_\alpha \psi \qquad \qquad \{\bar u W_\alpha \psi,\bar u\}_\BV = (-1)^{|\alpha|}\bar u W_\alpha \, ,
\end{equation}
generating a (graded) Lie algebra of world-valued $k \times k$ matrices. Including the action of $Q$
on words, this becomes the dg-Lie algebra $\fgl_k[\cA^!] \times \fgl_k[\cA^!]$.

The generators of ghost numbers $-1$ and $1$ associated to $\bar \psi W_\alpha \psi$ and $\bar u W_\alpha u$ respectively extend it to a dg-Lie algebra of world-valued $(k|k) \times (k|k)$ matrices $\fgl_{k|k}[\cA^!]$:
\begin{equation}
    \{\bar \psi W_\alpha \psi,u\}_\BV = W_\alpha \psi \qquad \qquad \{\bar \psi W_\alpha \psi,\bar u\}_\BV = - (-1)^{|\alpha|}\bar \psi W_\alpha \, ,
\end{equation}
and 
\begin{equation}
    \{\bar u W_\alpha u,\psi\}_\BV = (-1)^{|\alpha|}W_\alpha u \qquad \qquad \{\bar u W_\alpha u,\bar \psi\}_\BV = (-1)^{|\alpha|}\bar u W_\alpha \, .
\end{equation}
Finally, we need to account for the action of $\Delta_\BV$ mapping the $s\cM_\bullet$ to a multiple of $sT_\bullet$. We will denote the resulting dg-Lie algebra $\cL$ as $\fgl'_{k|k}[\cA^!]$. 

The second argument of the CE (co)homology complex conventionally denotes a representation of $\cL$, whose generators are contracted with the variables to give a linear term. Here the linear 
$sM^+_0 + sM^-_0$ denotes a 1-dimensional representation of the $\fgl_{k|k}$ zero-modes, the Berezinian $\Ber$. Hence the $\Ber^N$ argument.

A physicist would be slightly more familiar with the dual Chevalley-Eilenberg {\it cohomology} complex, under the name of BRST cohomology. In the case at hand, consider the following BV action
\begin{equation}
    S_\BV = \Tr_k m_0 \bar \psi \psi + \sum_{\alpha} \Tr_k \left[m_\alpha \bar \psi W_\alpha \psi + c_\alpha^+ \bar \psi W_\alpha u + c_\alpha^- \bar u W_\alpha \psi + r_\alpha \bar u W_\alpha u\right] + \sum_{[\alpha]} t_{[\alpha]} \Tr W_\alpha \, ,
\end{equation}
introducing couplings $t_\bullet$ dual to traces and $c_\bullet \equiv (m_\bullet, c^\pm_\bullet,r_\bullet)$ dual to mesons.

The action of $\Delta_\BV$ on $e^{S_\BV}$ pulls down mesons and single traces which can be traded for derivatives with respect to the couplings, allowing us to write  
\begin{equation}
    (Q + \Delta_\BV) \, e^{S_\BV} = \cQ \, e^{S_\BV} \, ,
\end{equation}
for a nilpotent differential operator $\cQ$. The action of $\cQ$ on $\bC[t_\bullet, c_\bullet]$ is the linear dual of the action of $Q + \Delta_\BV$ on $\mathbb{C}[T_\bullet,\cM_\bullet]$ and defines the cohomology complex   
\begin{equation}
    C^\bullet(\cL,\Ber^N) \equiv (\bC[t_\bullet, c_\bullet], \cQ)\, .
\end{equation} 
Explicitly, we have
\begin{equation}
    \cQ = Q^\vee + \sum_\alpha (-1)^{|\alpha|}\left( c_\alpha^- \cD_{-,\alpha} - c_\alpha^+\cD_{+,\alpha} \right )+ \cdots \, .
\end{equation}
and we recognize the formal structure of a BRST charge, with the open loop equations playing the role of gauge constraints with ghosts $c^\pm_\bullet$ and ghosts-for-ghosts $r_\bullet$ associated to the redundancy (\ref{eq:gaugegauge}). 

\subsection{Relatively Correct}

Landing in the well-understood environment of the CE (co)homology complex allows us to contemplate some useful variations of the problem. Recall our initial discussion about
the CE complex $C^\bullet(\fgl_N[\cA])$ and its relative version $C^\bullet(\fgl_N[\cA]|\fgl_N)\equiv \cH_{\cA;N}$: the ghost zeromode in the absolute complex $C^\bullet(\fgl_N[\cA])$ effectively imposes $U(N)$ invariance, but leaves behind extra $\Tr C^{2n+1}$ classes. 

Our original ``non-derived'' analysis certainly focused on a subset 
\begin{equation}
    \bC[T_\bullet, M_\bullet,M_0]^{U(k)_N\times U(k)_N}
\end{equation} of polynomials in a specific representation of $U(k) \times U(k)$. The CE complex 
\begin{equation}
    C^\bullet(\fgl'_{k|k}[\cA^!],\Ber^N) \, ,
\end{equation} 
instead, imposes that condition cohomologically with the help of the $c^\pm_0$ zeromodes. 

Thus it seems natural to consider a relative version of the complex, which would impose that constraint directly and drop the $c^\pm_0$ zeromodes. On the other hand, we certainly cannot drop the $m_0$ zeromode, which is an important ingredient of the non-derived construction and supports 
many cohomology classes of interest. Although in principle we could keep $r_0$, we will see momentarily that it would be unnatural. 

Accordingly, defining $\mathfrak{b}_{k|k}$ as the subalgebra generated by $sM_0^\pm$ and $sM^r_0$, we can consider the relative  
homology problem
\begin{equation}
    C_\bullet(\fgl_{k|k}[\cA^!]|\mathfrak{b}_{k|k};\Ber^N) \, . 
\end{equation}
Note that the relative complex sets $M_0^\pm=0$ and $M^r_0=0$ and quotients by the symmetries they generate, including in particular the action of $\{\bar u u,\bullet \}$. This is the shifted version of a Hamiltonian reduction. 

This complex is still natural from the perspective of studying the condition 
\begin{equation}
    Q P(T_\bullet,M_\bullet,M_0) + \Delta_\BV P^{(1)}(T_\bullet,\cM_\bullet,M_0) =0 \, .
\end{equation}
We simply decided that there is no point including $M^\pm_0$ in $P^{(1)}$, as $P$ was already transforming as $\det^N \times \det^N$. And at the next step we should then not include $M^r_0$ either, otherwise the bracket with $M_0$ will bring back $M^\pm_0$. We still want to impose the associated  gauge-for-gauge symmetry, but we do it by hand by taking a quotient. 

The usefulness of this choice of $\mathfrak{b}_{k|k}$-relative cohomology will be determined by the 
amount of spurious states which can obstruct the lift of $P$ to a class in $Q + \Delta_\BV$ cohomology. The integration map gives us a chain map from this cohomology to $\cH_{\cA;N}$, and the key question is how close to surjective and injective the map is. 
By construction, 
\begin{equation}
H_\bullet(\fgl_{k|k}|\mathfrak{b}_{k|k},\Ber^N) = \bC (\det M_0)^N
\end{equation}
has no spurious classes. Spurious classes do appear once $\cA$ is nontrivial, but experimentally they are pushed at higher and higher ghost number as $k$ is increased. In the next Section we will provide evidence that 
\begin{align}
    \lim_{k\rightarrow\infty} H_\bullet(\mathfrak{gl}_{k|k}'[\mathcal{A}^!]|\mathfrak{b}_{k|k},\Ber^N) \simeq \mathcal{H}_{\mathcal{A};N} \, .
\end{align}
We also run some cohomology computations which support this conjecture, see Appendix \ref{app:largek}. Furthermore, it appears that the absolute homology  
\begin{align}
    \lim_{k\rightarrow\infty} H_\bullet(\mathfrak{gl}_{k|k}'[\mathcal{A}^!],\Ber^N) \simeq \mathcal{H}_{\mathcal{A};N} \otimes \mathbb{C}[\sTr \cM_0,\sTr \cM_0^3,\dots]
\end{align}
also recovers the physical finite-$N$ cohomology dressed by some extra zero-mode classes.

\subsection{Higher Trace Maps}

We are now ready to give a D-brane interpretation to our results. As discussed before, $\cA^!$ is the algebra of open string modes on ``Koszul-dual'' D-branes which are fully transverse to the collection of D-branes associated to $\cA$.
The space parameterized by the $m_\bullet$ variables may be understood as parameterizing complexes of $k|k$ such D-branes. One way to understand this is that the open strings stretched between a single Koszul D-brane and the original stack of D-branes would only, say, give the analogue of a $\psi$ and $\bar u$ entry
in the Grassmann integral and would not lead to consistent answers: they would induce an anomaly on the original stack. Brane/antibrane pairs cancel the anomaly. 
In concrete examples, the Giant Graviton D-branes used to describe local operators are indeed typically of co-dimension 1 in the transverse space and can be thought of as the result of tachyon condensation on Koszul-dual brane/antibrane pairs. 

This motivates a simple proposal: $C^\bullet(\fgl_{k|k}[\cA^!]|\mathfrak{b}_{k|k},\Ber^N)$ is the natural geometric quantization of the space of complexes of $k|k$ Koszul-dual D-branes, accounting for the gauge symmetries of the complex. Our map $C_\bullet(\fgl_{k|k}[\cA^!]|\mathfrak{b}_{k|k},\Ber^N) \to \cH_{\cA;N}$
describes how (the dual of) this space of Giant Graviton states maps to the $Q$-cohomology of local operators of the $U(N)$ gauge theory. It appears that the Giant Graviton quantization at any finite $k$ includes spurious states which are killed by the map to the gauge theory, and may also potentially miss some states in $\cH_{\cA;N}$, but the problem is ameliorated by increasing $k$. 

These formal considerations can be made much more concrete in B-model examples by simplifying the complexes $\cA^!$ and $CC_r^\bullet(\cA)$ to their cohomology $H\!\cA^!$ and $HC_r^\bullet(\cA)$
which have a geometric interpretation. 

The absolute CE complex is defined for $L_\infty$ algebras and its cohomology is invariant under quasi-isomorphism of $L_\infty$ algebras, e.g. polynomial redefinitions of the elementary variables which are a quasi-isomorphism at the linear level. The relative CE complex behaves well as long as the quasi-isomorphism is $\mathfrak{b}_{k|k}$-equivariant. 

In practice, that means we are potentially allowed to simplify $\fgl'_{k|k}[\cA^!]$ to a simpler $L_\infty$ algebra $\fgl'_{k|k}[H\!\cA^!]$, where $H\!\cA^!$ denotes the cohomology of the dg-algebra $\cA^!$. Then the new $\cM_\bullet$ variables are valued in $H\!\cA^!$ and the new $T_\bullet$ are valued in $HC_r^\bullet(\cA)$. In doing so, we need to insure a quasi-isomorphism for the whole $L_\infty$ algebra and in particular transport appropriately the $A_\infty$ structure of $\cA^!$ and the coupling between $\cA^!$ and $CC_r^\bullet(\cA)$ given by the trace map 
\begin{equation}
    W_\alpha \to \Tr W_\alpha \, .
\end{equation}
It is well-known that $H\!\cA^!$ can be equipped with an $A_\infty$ structure which makes it quasi-isomorphic to $\cA^!$. That also makes $\fgl_{k|k}[\cA^!]$ quasi-isomorphic to $\fgl_{k|k}[H\!\cA^!]$. The trace map is a bit more subtle. 

The trace map certainly commutes with the action of $Q$, e.g. gives a chain map $\cA^! \to CC^{\bullet-1}_r(\cA)$. In string field theory language, it functions as an open-closed vertex mapping an open string to the closed string state created by a $\cD^!$ boundary with the open string insertion. A powerful way to understand the trace map is that it defines a 
perfect pairing 
\begin{equation}
    HC_r^\bullet(\cA) \otimes HC_r^{\bullet-1}(\cA^!) \to \bC
\end{equation}
between the cyclic cohomologies of $\cA$ and $\cA^!$, supported on $HC_r^0(\cA^!)$, aka linear functions on $\cA^!$. The string theory logic of this statement is that both $\cA$ and $\cA^!$ are open string algebras of D-branes in the same theory of closed strings, so they should carry compatible information about them. 

The power of this statement is that quasi-isomorphisms of $A_\infty$ algebras lift combinatorially to quasi-isomorphisms of their cyclic cohomology, so $HC_r^{\bullet-1}(\cA^!) \simeq HC_r^{\bullet-1}(H\!\cA^!)$. The trace map thus maps a perfect pairing 
\begin{equation}
    HC_r^\bullet(\cA) \otimes HC_r^{\bullet-1}(H\!\cA^!) \to \bC
\end{equation}
as well. The only subtlety is that the pairing now involves both the naive trace and ``higher trace maps'' $\Tr_n$ which take a cyclic sequence of $n$ elements in $H\!\cA^!$ to a class in $HC_r^{\bullet-n}(\cA)$. 

Because of the perfect pairing, we do not even need to think about the higher trace maps. We can just take the new $T_\bullet$ to be valued in the dual to $HC_r^{\bullet-1}(H\!\cA^!)$ instead and forget about 
$HC_r^\bullet(\cA)$ altogether. All these considerations and the higher trace maps can be extended to matrices valued in $\cA^!$ as well, so that these $T_\bullet$ also come with a collection of 
higher trace maps taking cyclic sequences of elements of $\fgl_{k|k}[H\!\cA^!]$ to $T_\bullet$. 

After the dust settles, we find a natural candidate definition for the $L_\infty$ algebra $\fgl'_{k|k}[H\!\cA^!]$ and the associated CE homology complex: we take generators $M_\bullet$ and $T_\bullet$ in $\fgl_{k|k}[H\!\cA^!]$ and in the dual to $HC_r^{\bullet}(H\!\cA^!)$ 
and include in the differential both the $\fgl_{k|k}[H\!\cA^!]$ multiplication and the higher trace maps, symmetrized to act on polynomials of $M_\bullet$
to produce $T_\bullet$. 

This definition of the $L_\infty$ algebras $\fgl'_{k|k}[{\mathcal B}]$ makes sense for any $A_\infty$ algebra ${\mathcal B}$, and it would be interesting to explore it further. For now, we will just assume a key conjecture: 
\begin{mdframed}[
    linecolor=none,
    tikzsetting={draw=black,dashed,line width=0.4pt},
    innertopmargin=10pt,
    innerbottommargin=15pt
]
\begin{equation} \label{LkIso}
    \fgl'_{k|k}[H\!\cA^!] \simeq \fgl'_{k|k}[\cA^!]
\end{equation}\end{mdframed}
and thus 
\begin{mdframed}[
    linecolor=none,
    tikzsetting={draw=black,dashed,line width=0.4pt},
    innertopmargin=10pt,
    innerbottommargin=15pt
]
\begin{equation}
C^\bullet(\fgl_{k|k}'[H\!\cA^!]|\mathfrak{b}_{k|k},\Ber^N) \simeq C^\bullet(\fgl_{k|k}'[\cA^!]|\mathfrak{b}_{k|k},\Ber^N) \, ,
\end{equation}\end{mdframed}
which manifestly reduces the complex to something which only employs intrinsic properties of the Koszul dual D-branes and, in B-model examples, the geometry of Giant Graviton D-branes.

\section{Categorification of Murthy's Giant Graviton Expansion}
\label{sec:MGGE}

In the previous Section, we introduced complexes that impose the open loop equations cohomologically:
\begin{align}
H^\bullet (\mathfrak{gl}_{k|k}'[\cA^!]|\mathfrak{b}_{k|k},\Ber^N) \, .\label{eq:complexb}
\end{align}
These cohomologies provide candidate description of the (dual) finite-$N$ physical cohomology $\cH_{\cA;N}^\vee$, but can contain additional spurious classes at finite $k$.

In this section, we compare their graded characters with Murthy's Giant Graviton Expansion formulas \cite{Murthy:2022ien, Liu:2022olj, Eniceicu:2023uvd, Chen:2024cvf}. We conjecture that the character of \eqref{eq:complexb} is the partial
MGGE sum up to the $k$th correction, multiplied by the large-$N$
index. We test this conjecture for the algebras $\cA=\mathbb C[\theta]$
and $\cA=\mathbb C[z]$. Murthy's formula then implies that these characters approach
the finite-$N$ physical index as $k\to\infty$. We take this
as evidence that the spurious classes disappear in this limit and conjecture
\begin{align}
\lim_{k\rightarrow\infty} H^\bullet(\mathfrak{gl}'_{k|k}[\cA^!]|\mathfrak{b}_{k|k},\Ber^N) \simeq \mathcal{H}_{\cA;N}^\vee \, . \label{eq:repeated}
\end{align}

\subsection{Murthy's Giant Graviton Expansion}

Recall that the graded character $Z_N=\chi(\mathcal{H}_{\cA;N})$ is computed by the standard Molien-Weyl $U(N)$ matrix integral. Including fundamental and antifundamental variables, we have
\begin{equation}
    Z_N[f,\bar{v},v] = \frac{1}{N!} \oint_{|\mu_i|=1} \prod_{i=1}^N \frac{d \mu_i}{2 \pi \i \mu_i} \prod_{i\neq j}^N \qty(1-\frac{\mu_i}{\mu_j} )\PE\qty[ f \sum_{i,j=1}^N \frac{\mu_i}{\mu_j} + v \sum_{i=1}^N \mu_i + \bar{v} \sum_{i=1}^N \frac{1}{\mu_i} ] \, .
\end{equation}
Here $f$, $v$, and $\bar v$ are the single-letter indices of the adjoint, fundamental, and antifundamental variables, respectively. In the adjoint case, $(1-f)$ is the graded character of $\cA$. The plethystic exponential $\PE$ computes the equivariant character of a space of polynomials (aka a Fock space) given the equivariant character of the space of variables (aka single-particle states).

These generating functions have a good large $N$ (aka planar) limit 
\begin{equation}
    Z_\infty[f,\bar{v},v] = \frac{\PE\left[\frac{\bar v v}{1-f}\right]}{\prod_{k=1}^\infty (1-f_k)} \, ,
\end{equation} 
with $f_k$ being $f$ with all arguments/fugacities raised to the power of $k$. This counts polynomials in single-trace operators and mesons. The difference $Z_\infty - Z_N$ is due to trace relations, and thus involves monomials in the fugacities with degree of order $N$. 

Murthy's GGE \cite{Murthy:2022ien} applies very generally to fee partition functions/indices $Z_N[f,\bar v,v]$ and is (to the best of our knowledge), the only proven GGE formula. The proof relates  
$Z_N[f,\bar v,v]$ to $Z_N[0,\bar v,v]$ by a Hubbard-Stratonovich transformation
and then derives a GGE for $Z_N[0, \bar v,v]$ by combinatorial means. It takes the form  
\begin{equation}
    Z_N = Z_\infty\sum_{k=0}^\infty G_{N}^{(k)} \, . \label{eq:MGGE}
\end{equation}
The $N$-dependent corrections $G_{N}^{(k)}$ can be recast as contour integrals \cite{Eniceicu:2023uvd,Liu:2022olj, Chen:2024cvf}: 
\begin{align}
G_N^{(k)}[f,\bar v,v]
={}&
\frac{1}{(k!)^2}
\oint_{\substack{|\mu_i|=1-\epsilon\\|\nu_i|=1+\epsilon}}
\prod_{i=1}^{k}
\qty[
\frac{d\mu_i}{2\pi\i \mu_i}
\frac{d\nu_i}{2\pi\i \nu_i}
\qty(\frac{\mu_i}{\nu_i})^N ] \frac{
\prod_{i\neq j}^{k}
\qty(1-\frac{\mu_i}{\mu_j})\qty(1-\frac{\nu_i}{\nu_j}) }{ \prod_{i,j=1}^{k} \qty(1-\frac{\mu_i}{\nu_j})\qty(1-\frac{\nu_j}{\mu_i})}
\nonumber \\
&\times\PE\Biggl[
-\frac{f}{1-f}
\sum_{i,j=1}^{k}
\qty(\mu_i-\nu_i)\qty(\frac{1}{\mu_j}-\frac{1}{\nu_j})  \label{eq:GkN} \\
&\hspace{6em}
+\frac{\bar v}{1-f}\sum_{i=1}^{k}(\mu_i-\nu_i)
-\frac{v}{1-f}\sum_{i=1}^{k}\qty(\frac{1}{\mu_i}-\frac{1}{\nu_i})
\Biggr] \, . \nonumber
\end{align}
The integration contours are chosen so that both $(1-\mu_i /\nu_j)$ and $(1-\nu_j / \mu_i)$ denominator factors can be expanded in powers of $\mu_i / \nu_j$. 

The integral \eqref{eq:GkN} resembles a Molien-Weyl integral for $U(k|k)$ supergroup, for some collection of adjoint and (anti)fundamental letters and an insertion of the $N$th power of the Berezinian (superdeterminant) \cite{Eniceicu:2023uvd}. Below, we review the formulas for supergroup invariants and conjecture an extension to cohomology with Berezinian coefficients, relative to the constant block-lower-triangular subalgebra $\mathfrak b_{k|k}$. We conjecture that the characters of this cohomology are precisely those appearing in Murthy's GGE.

\subsection{Supergroup Invariants and Koszul-dual D-branes}

The analogue of the Molien-Weyl integral for supergroups $U(n+K|K)$ was recently proposed by one of the authors in \cite{Budzik:2025vzr}, and proved for
adjoint variables in \cite{berele2025invariants}. The claim is that $U(n+K|K)$ invariants built from adjoint and (anti)fundamental variables are counted by a sum 
\begin{equation}
    Z_{n+K|K}=\sum_{k=0}^K Z_{n+k|k}^+ \, , \label{eq:psum}
\end{equation}
where $Z_{n+k|k}^+$ is computed by a contour integral over the Cartan of $U(n+k|k)$, with an integrand which is the obvious supergroup extension of the Molien-Weyl integrand
\begin{align}
Z^+_{n|m}\qty[f,\bar v,v]
={}&
\frac{1}{n!m!}
\oint_{\substack{\qty|\mu_i|=1-\epsilon\\\qty|\nu_a|=1+\epsilon}}
\prod_{i=1}^{n}\frac{\dd\mu_i}{2\pi i\mu_i} \prod_{a=1}^{m}\frac{\dd\nu_a}{2\pi i\nu_a}
\frac{
\prod_{i\neq j}^{n}\qty(1-\frac{\mu_i}{\mu_j})
\prod_{a\neq b}^{m}\qty(1-\frac{\nu_a}{\nu_b})
}{
\prod_{i=1}^{n}\prod_{a=1}^{m}
\qty(1-\frac{\mu_i}{\nu_a})\qty(1-\frac{\nu_a}{\mu_i})
}
\nonumber\\
&\times\PE\qty[
f\qty(\sum_{i=1}^{n}\mu_i-\sum_{a=1}^{m}\nu_a)
\qty(\sum_{i=1}^{n}\frac{1}{\mu_i}-\sum_{a=1}^{m}\frac{1}{\nu_a})
] \label{eq:Zplus} \\
&\times\PE\qty[
v\qty(\sum_{i=1}^{n}\mu_i-\sum_{a=1}^{m}\nu_a)
+\bar v\qty(\sum_{i=1}^{n}\frac{1}{\mu_i}-\sum_{a=1}^{m}\frac{1}{\nu_a})
]. \nonumber
\end{align}

At $N=0$, the Berezinian factors in MGGE corrections \eqref{eq:GkN} are absent. The remaining integrand has precisely the form \eqref{eq:Zplus}, with transformed single-letter indices. It was observed in \cite{Budzik:2025vzr}, the single-letter index appearing in \eqref{eq:GkN} is that associated with $\cA^!$. Writing the graded characters of $\cA$ and $\cA^!$ as $1-f$ and $1-f^!$, respectively, they satisfy the relation\footnote{A matrix version of this relation appears in the quiver generalization of Murthy's GGE  \cite{Purkayastha:2025whi} and was used in \cite{Budzik:2026vqe}.}
\begin{align}
\qty(1-f)\qty(1-f^!)=1 \, ,
\qquad f^!=-\frac{f}{1-f} \, .
\label{eq:koszul_index}
\end{align}
We can also extend it to (anti)fundamental variables by
\begin{align}
    v^! \equiv \frac{\bar{v}}{1-f} \, , \qquad\qquad \bar{v}^! \equiv -\frac{v}{1-f} \, .
\end{align}
Therefore, at $N=0$, the corrections $G^{(k)}_0$ in Murthy's GGE are the supergroup contour integrals evaluated with the Koszul-dual single-letter indices
\begin{align}
G_0^{(k)}\qty[f,\bar v,v]=Z^+_{k|k}\qty[f^!,\bar v^!,v^!] \, .
\end{align}

For certain superconformal indices, a B-model holographic
interpretation of this relation was proposed in \cite{Budzik:2025vzr} (extended to thermal free partition functions in \cite{Budzik:2026vqe}). The algebra $\cA^!$ describes open strings on Koszul-dual D-branes transverse to the original stack. Its worldvolume theory has gauge supergroup $U(K|K)$ and adjoint single-letter index $f^!$. The partial MGGE sum is then interpreted as the graded partition function of a system of $K$ such branes and $K$ antibranes. 

Below we discuss the integrals  \eqref{eq:Zplus} with the extra Berezinian insertion.

\subsection{Berezinian Coefficients}

For $N>0$, the integrals \eqref{eq:GkN} contain extra Berezinian factors \cite{Eniceicu:2023uvd}:
\begin{align}
\Ber\qty(
\operatorname{diag}\qty(\mu_1,\ldots,\mu_k | \nu_1,\ldots,\nu_k))^N = \prod_{i=1}^k\qty(\frac{\mu_i}{\nu_i})^N \, .
\end{align}

We therefore want to study the generalization of the contour integrals \eqref{eq:Zplus} by insertion of the $N$th power of the Berezinian. In this Section, we specialize to equal ranks and adjoint variables $\bar v= v=0$:
\begin{align}
\begin{split}
Z^+_{k|k}\qty[f;N]
={}
\frac{1}{(k!)^2}
\oint_{\substack{\qty|\mu_i|=1-\epsilon\\\qty|\nu_i|=1+\epsilon}}
\prod_{i=1}^{k} & \qty[\frac{\dd\mu_i}{2\pi i\mu_i}\frac{\dd\nu_i}{2\pi i\nu_i}
\qty(\frac{\mu_i}{\nu_i})^N ]
\frac{
\prod_{i\neq j}^{k}\qty(1-\frac{\mu_i}{\mu_j})\qty(1-\frac{\nu_i}{\nu_j})
}{
\prod_{i,j=1}^{k}
\qty(1-\frac{\mu_i}{\nu_j})\qty(1-\frac{\nu_j}{\mu_i})
} \\
&\times\PE\qty[
f\sum_{i,j=1}^{k}
\qty(\mu_i-\nu_i)\qty(\frac{1}{\mu_j}-\frac{1}{\nu_j})
] \, .
\end{split} \label{eq:Zplus_Ber} 
\end{align} 

We conjecture that the partial sums of integrals with the Berezinian factor \eqref{eq:Zplus_Ber} compute the graded character of cohomology relative to the constant block-lower-triangular subalgebra $\mathfrak b_{k|k}$, with coefficients in $\Ber^N$:
\begin{mdframed}[
    linecolor=none,
    tikzsetting={draw=black,dashed,line width=0.4pt},
    innertopmargin=3pt,
    innerbottommargin=15pt
]\begin{align}
\chi(H^\bullet(\mathfrak{gl}_{K|K}[\cA] | \mathfrak{b}_{K|K},\Ber^N) = \sum_{k=0}^K Z^+_{k|k}[f;N] \, , \label{eq:Ab}
\end{align}
\end{mdframed}
where $1-f$ is the graded character of $\cA$. For $N=0$, the right hand side reduces to the partial sum computing $U(K|K)$ supergroup invariants \eqref{eq:psum}. This is consistent as we expect
\begin{align}
\chi(H^\bullet(\mathfrak{gl}_{K|K}[\cA] | \mathfrak{gl}_{K|K}) = \chi(H^\bullet(\mathfrak{gl}_{K|K}[\cA] | \mathfrak{b}_{K|K}) \, . \label{eq:expected}
\end{align}

We also find experimentally that
\begin{mdframed}[
    linecolor=none,
    tikzsetting={draw=black,dashed,line width=0.4pt},
    innertopmargin=6pt,
    innerbottommargin=18pt
]\begin{align}
\chi(H^\bullet(\mathfrak{gl}_{k|k}[\cA] | \mathfrak{gl}_{k|k},\Ber^N) = Z^{+}_{k|k}[f;N] \,, \qquad N>0 \, . \label{eq:Agl}
\end{align}
\end{mdframed}
We test both conjectures for examples $\cA=\mathbb{C}[\theta]$ and $\cA=\mathbb{C}[z]$ in Appendix \ref{app:relative}.

For unequal ranks and $N>0$, both relative cohomologies considered above vanish. This is because for $n\neq m$ both $\mathfrak{gl}_{n|m}$ and $\mathfrak{b}_{n|m}$ contain the identity $\mathbf{1}_{n|m}$, and there are no non-zero cochains valued in $\Ber^N$ invariant under $\mathbf{1}_{n|m}$ for $N>0$. This matches the checks that the contour integrals $Z^+_{n|m}[f;N]$ vanish for $n\neq m$ and $N>0$ in all examples tested. 

It would be interesting to understand the Berezinian factor in the integrals as counting the BV resolution of the $\Ber$ representation, but we leave that for future work.

\subsection{Large-$k$ Limit}

We recognize that the partial sums in MGGE \eqref{eq:GkN} count the cohomology relative to the constant block-lower-triangular subalgebra $\mathfrak b_{k|k}$, with coefficients in $\Ber^N$ for the algebra $\cA^!$, with graded character $1-f^!$. Including the contribution from the $t_\bullet$, which are counted by $Z_\infty$, the character of $C^\bullet(\fgl'_{k|k}[\cA^!]|\mathfrak{b}_{k|k},\Ber^N)$ is thus identified with 
\begin{equation}
    Z_\infty \sum_{k=0}^K G_{N}^{(k)}
\end{equation}
and in the $K \to \infty$ limit reproduces the Murthy's formula \eqref{eq:MGGE} for $Z_N$.

This therefore leads us to conjecture that the spurious classes drop out at large $k$ and
\begin{mdframed}[
    linecolor=none,
    tikzsetting={draw=black,dashed,line width=0.4pt},
    innertopmargin=8pt,
    innerbottommargin=18pt
]\begin{align}
    \lim_{k\rightarrow\infty} H^\bullet(\mathfrak{gl}_{k|k}'[\cA^!]|\mathfrak{b}_{k|k}, \Ber^N) \simeq \cH_{\cA;N}^\vee \, . \label{eq:largek}
\end{align}\end{mdframed}
We test this conjecture for $\cA=\bC[\theta]$ and $\cA=\bC[z]$ in Appendix \ref{app:largek}.

In conclusion, our formalism ``categorifies'' Murthy's GGE to the quantization of a moduli space of Giant Graviton D-branes. 

One could also consider the absolute cohomology or other relative cohomologies that would include extra classes containing various zero-modes. In particular, we conjecture that the absolute cohomology recovers $\cH_{\cA;N}^\vee$ dressed by extra ghost classes:
\begin{align}
    \lim_{k\rightarrow\infty} H^\bullet(\mathfrak{gl}_{k|k}'[\cA^!], \Ber^N) \simeq \cH_{\cA;N}^\vee \otimes\mathbb{C}[\sTr c_0 \, , \sTr c_0^3 , \dots] \, .
\end{align}

Considering the cohomology relative with respect to $\mathfrak{gl}_k\oplus\mathfrak{gl}_k\subset\mathfrak{gl}_{k|k}$ will leave $r_0$ zero-modes. We therefore expect:
\begin{align}
    \lim_{k\rightarrow\infty} H^\bullet(\mathfrak{gl}_{k|k}'[\cA^!]|\mathfrak{gl}_k\oplus\mathfrak{gl}_k, \Ber^N) \simeq \cH_{\cA;N}^\vee \otimes\mathbb{C}[\Tr(m_0r_0) , \Tr(m_0r_0)^2 , \dots ].
\end{align}

\acknowledgments

We would like to thank Kevin Costello, Elliott Gesteau, Matthew Heydeman, Ji Hoon Lee, Dan Kapec, Jacob McNamara, Atul Sharma and Edward Witten for helpful conversations. The work of KB is supported by the Simons Collaboration on Celestial Holography. This
research was supported in part by a grant from the Krembil Foundation. DG is
supported by the NSERC Discovery Grant program and by the Perimeter Institute
for Theoretical Physics. Research at Perimeter Institute is supported in part by
the Government of Canada through the Department of Innovation, Science and
Economic Development Canada and by the Province of Ontario through the Ministry
of Colleges and Universities. The conjectures were produced by humans and tested using Python code produced by ChatGPT and Claude. AI tools were also employed for minor text editing.

\appendix

\section{Examples of Open Loop Equations}
\label{app:SD}

In this Appendix, we illustrate the module presentation and OLE/SD equations for the space of invariants $\mathcal{H}_{d;N}\simeq \mathcal{R}_{\leq}$.

\subsection{$d=1$}
\label{subsubsec:TRd1}

For a single matrix $X$, traces $T_k$ and words $W_k$ can be labeled by positive integers $k\in\mathbb{Z}_+$,
and the concatenation of words corresponds to the addition of integers, i.e. $W_{k_1} W_{k_2}=W_{k_1+k_2}$. The multi-indices are tuples of positive integers $A=(k_1,\dots,k_n)$.

The SD equations \eqref{eq:role} both take the form
\begin{equation}
    T_{n} R_{(n_1, \cdots, k_n)} = R_{(k,k_1, \cdots, k_n)} + \sum_{i=1}^n R_{(k_1, \dots , k_i+k , \dots , k_n)} \, .
\end{equation}
We can use the above equations to eliminate all factors of $T_{k}$ in $\cR_\leq$, reducing it to $\oplus_{|A|\leq N} R_A$. Looking at summands with fixed total number of $X$, we immediately match $\cH_{1;N}$ dimension-wise: generators $R_{|A|}$ with $|A|\leq N$ are labeled by Young Tableaux of width up to $N$, while monomials in $T_k$ with $k\leq N$ are labeled by Young Tableaux of height up to $N$.

We can apply the same considerations to $\cR_{>}$, reducing it to $\oplus_{|A|> N} R_A$ to see a dimension match with ${\cal T}_{1;N}$: both are associated to Young Tableaux of width/height greater than $N$.

\subsection{$d=2$}

Now we give some examples of the OLE equations in the two-matrix case.

We will denote the two matrices by $X,Y$. Their associated $U(1)$ charges count the numbers of $X$ and $Y$, respectively. The traces  $T_\bullet$ are labeled by cyclic words and the generators $R_\bullet$ by an unordered collection of ordered words, e.g.
\begin{align}
    T_{[XY]}=T_{[YX]} \, , \qquad R_{(X,Y)}=R_{(Y,X)} \, .
\end{align}

The simplest SD equations involving both matrices $X,Y$ occur in the $(1,1)$ charge sector. Before imposing SD equations, this sector has 7 basis elements:
\begin{gather}
    T_{[XY]}R_{\emptyset} \, , \quad T_{[X]} T_{[Y]} R_\emptyset \nonumber \\
    T_{[Y]} R_{(X)} \, , \quad T_{[X]} R_{(Y)} \label{eq:d2gen1} \\
    R_{(XY)} \, , \quad R_{(YX)} \, , \quad R_{(X,Y)} \, . \nonumber
\end{gather}

The zero-slot SD equations are
\begin{equation}
\begin{split}
    T_{[XY]} R_\emptyset &= R_{(XY)} \\
    T_{[X]} T_{[Y]} R_\emptyset &= T_{[X]}  R_{(Y)} \\
    T_{[X]} T_{[Y]} R_\emptyset &= T_{[Y]}  R_{(X)} \, ,
\end{split} \label{eq:d2sd1}
\end{equation}
where the last two equations are obtained by multiplying charge $(1,0)$ and $(0,1)$ zero-slot SD equations by a trace. The one-slot SD equations are
\begin{equation}
\begin{split}
    T_{[Y]} R_{(X)} &= R_{(X,Y)} + R_{(XY)} \\
    T_{[Y]} R_{(X)} &= R_{(X,Y)} + R_{(YX)} \\
    T_{[X]} R_{(Y)} &= R_{(X,Y)} + R_{(YX)} \\
    T_{[X]} R_{(Y)} &= R_{(X,Y)} + R_{(XY)} \, .
\end{split} \label{eq:d2sd2}
\end{equation}

Before imposing $N$ cutoff, the SD equations \eqref{eq:d2sd1} and \eqref{eq:d2sd2} have rank five. For $N\geq 2$, the cutoff imposes no further relations, so the SD equations reduce the number of generators \eqref{eq:d2gen1} to two, e.g.
\begin{align}
    T_{[XY]} R_\emptyset \, , \qquad  T_{[X]}T_{[Y]} R_\emptyset \, ,
\end{align}
which map to $\Tr XY$, $\Tr X\Tr Y$.

For $N=1$, the cutoff imposes $R_{(X,Y)}=0$. The SD equations then leave one generator, e.g.
\begin{align}
    T_{[X]}T_{[Y]} R_\emptyset \, .
\end{align}
In particular, combining the SD equations \eqref{eq:d2sd1} and \eqref{eq:d2sd2}, gives
\begin{align}
    \qty( T_{[X]} T_{[Y]} - T_{[XY]} ) R_\emptyset = R_{(X,Y)} = 0 \,.
\end{align}
Therefore, after setting the $R$-generator with $N+1$ slots to zero and expressing the other $R$-generators using traces, the boundary SD equations recover the trace relation at $N=1$:
\begin{align}
    \Tr X \Tr Y - \Tr XY = 0 \, .
\end{align}

\subsection{$d=3$}

Now we give some examples of SD equations for three matrices $X,Y,Z$.

The simplest SD equations involving each matrix $X,Y,Z$ occur in the $(1,1,1)$ charge sector. This sector has 34 basis elements before imposing $N$ cutoff:
\begin{gather}
T_{[XYZ]}R_{\emptyset} \, , \quad
T_{[XZY]}R_{\emptyset} \, , \quad T_{[XY]}T_{[Z]}R_{\emptyset} \, , \quad
T_{[XZ]}T_{[Y]}R_{\emptyset} \, , \quad
T_{[YZ]}T_{[X]}R_{\emptyset} \, , \quad
T_{[X]}T_{[Y]}T_{[Z]}R_{\emptyset}
\nonumber \\
T_{[YZ]}R_{(X)} \, , \quad
T_{[XZ]}R_{(Y)} \, , \quad
T_{[XY]}R_{(Z)} \, , \quad
T_{[Y]}T_{[Z]}R_{(X)} \, , \quad
T_{[X]}T_{[Z]}R_{(Y)} \, , \quad
T_{[X]}T_{[Y]}R_{(Z)}
\nonumber \\
T_{[Z]}R_{(XY)} \, , \quad
T_{[Z]}R_{(YX)} \, , \quad
T_{[Y]}R_{(XZ)} \, , \quad
T_{[Y]}R_{(ZX)} \, , \quad
T_{[X]}R_{(YZ)} \, , \quad
T_{[X]}R_{(ZY)}
\nonumber \\
R_{(XYZ)} \, , \quad R_{(XZY)} \, , \quad
R_{(YXZ)} \, , \quad R_{(YZX)} \, , \quad
R_{(ZXY)} \, , \quad R_{(ZYX)} \label{eq:d3gen1} \\
T_{[Z]}R_{(X,Y)} \, , \quad
T_{[Y]}R_{(X,Z)} \, , \quad
T_{[X]}R_{(Y,Z)}
\nonumber \\
R_{(XY,Z)} \, , \quad R_{(YX,Z)} \, , \quad
R_{(XZ,Y)} \, , \quad R_{(ZX,Y)} \, , \quad
R_{(YZ,X)} \, , \quad R_{(ZY,X)}
\nonumber \\
R_{(X,Y,Z)} \, .
\nonumber
\end{gather}

The total number of SD equations in this charge sector is 47, including SD equations multiplied by traces. For $N\geq 3$, they have rank 28 and can be used to reduce \eqref{eq:d3gen1} to six representatives
\begin{gather}
    T_{[XYZ]}R_{\emptyset} \, , \quad
T_{[XZY]}R_{\emptyset} \, , \\
T_{[XY]}T_{[Z]}R_{\emptyset} \, , \quad
T_{[XZ]}T_{[Y]}R_{\emptyset} \, , \quad
T_{[YZ]}T_{[X]}R_{\emptyset} \, , \quad
T_{[X]}T_{[Y]}T_{[Z]}R_{\emptyset} \, ,
\end{gather}
which map to $\Tr XYZ$, $\Tr XZY$, $\Tr XY\Tr Z$, $\Tr YZ\Tr X$, $\Tr XZ\Tr Y$, $\Tr X\Tr Y\Tr Z$.

For $N=2$, the cutoff sets $R_{(X,Y,Z)}=0$. The SD equations can be combined to give
\begin{align}
    \left( T_{[X]}T_{[Y]}T_{[Z]} -T_{[X]}T_{[YZ]} -T_{[Y]}T_{[XZ]} -T_{[Z]}T_{[XY]} +T_{[XYZ]}+T_{[XZY]} \right) R_\emptyset &=R_{(X,Y,Z)} \, , \label{eq:d3tracerel}
\end{align}
which reproduces the $N=2$ trace relation:
\begin{align}
    \Tr X \Tr Y \Tr Z - \Tr X\Tr YZ -\Tr Y \Tr XZ - \Tr Z \Tr XY +\Tr XYZ + \Tr XZY &=0 \, .
\end{align}

For $N=1$, the cutoff imposes $R_{A}=0$ for $|A|\geq 2$. The SD equations can be combined to 
\begin{align}
    \bigl(T_{[X]}T_{[Y]}T_{[Z]}-T_{[X]}T_{[YZ]}\bigr)R_\emptyset
    &=T_{[X]}R_{(Y,Z)} = 0 \, ,\\
    \bigl(T_{[X]}T_{[Y]}T_{[Z]}-T_{[Y]}T_{[XZ]}\bigr)R_\emptyset
    &=T_{[Y]}R_{(X,Z)} = 0\, ,\\
    \bigl(T_{[X]}T_{[Y]}T_{[Z]}-T_{[Z]}T_{[XY]}\bigr)R_\emptyset
    &=T_{[Z]}R_{(X,Y)} =0 \, ,\\
    \bigl(T_{[X]}T_{[YZ]}-T_{[XYZ]}\bigr)R_\emptyset
    &=R_{(YZ,X)} = 0 \, ,\\
    \bigl(T_{[Y]}T_{[XZ]}-T_{[XZY]}\bigr)R_\emptyset
    &=R_{(XZ,Y)} = 0 \, ,
\end{align}
which recover the $N=1$ trace relations. This leaves one generator, e.g. $T_{[X]}T_{[Y]}T_{[Z]}R_\emptyset$, which maps to $XYZ$. 

\section{$\mathfrak{b}_{k|k}$ and $\mathfrak{gl}_{k|k}$ Relative Cohomology}
\label{app:relative}

In this Appendix we include tests of conjectures \eqref{eq:Ab} and \eqref{eq:Agl} for $\cA=\bC[\theta]$ and $\cA=\bC[z]$.
We compute the cohomology at finite $k$ using Python code produced by ChatGPT and Claude, which we intend to include in version 2, with more examples.

\subsection{$\cA=\mathbb{C}[\theta]$}
\label{subsec:1Ctheta}

First we consider the case of one bosonic matrix, i.e. $\cA=\mathbb{C}[\theta]$ with the single letter index $f=x$.

\paragraph{\underline{$N=0$}} For $N=0$, both the $\mathfrak b_{k|k}$-relative and the $\mathfrak{gl}_{k|k}$-relative cohomology agree as expected in \eqref{eq:expected}. They are concentrated in ghost number zero and their character is:
\begin{align}
    \chi(H^0(\mathfrak{gl}_{1|1}[\mathbb{C}[\theta]]|\mathfrak{b}_{1|1})) &= \chi(H^0(\mathfrak{gl}_{1|1}[\mathbb{C}[\theta]]|\mathfrak{gl}_{1|1})) \\
    &= 1 + x + 2x^2 + 3x^3 + 4x^4 + 5x^5 \\
    &\quad + 6x^6 + 7 x^7 + 8x^8 + 9x^9 + \mathcal{O}(x^{10}) \\
    \chi(H^0(\mathfrak{gl}_{2|2}[\mathbb{C}[\theta]]|\mathfrak{b}_{2|2})) &= \chi(H^0(\mathfrak{gl}_{2|2}[\mathbb{C}[\theta]]|\mathfrak{gl}_{2|2})) \\
    &= 1 + x + 2x^2 + 3x^3 + 5x^4 + 7x^5 \\
    &\quad + 11x^6 + 15x^7 + 22x^8 + 29x^9 + \mathcal{O}(x^{10}) \, .
\end{align}
The corresponding contour integrals were computed in \cite{Budzik:2025vzr}:
\begin{align}
    Z^+_{k|k}[x;0] = Z^+_{k|k}[x]= \frac{x^{k^2}}{\prod_{i=1}^k(1-x^i)^2} \, .
\end{align}
Their partial sum are
\begin{align}
    1+Z_{1|1}[x] &= 1 + x + 2 x^2 + 3 x^3 + 4 x^4 + 5 x^5 \\
    &\quad + 6 x^6 + 7 x^7 + 8 x^8 + 9 x^9 + \mathcal{O}(x^{10}) \\ 
    1+Z_{1|1}^+[x] +Z^+_{2|2}[x] &= 1 + x + 2 x^2 + 3 x^3 + 5 x^4 + 7 x^5 \\
    &\quad + 11 x^6 + 15 x^7 + 22 x^8 + 29 x^9 + \mathcal{O}(x^{10}) \, . 
\end{align}
These agree with the characters of the relative cohomologies.

\paragraph{\underline{$N=1$}}

For $N=1$, the computed $\mathfrak b_{k|k}$-relative cohomology contains only one class in ghost number zero corresponding to $\det m_0$. This was checked through $x^{11}$ for $k=1$ and through $x^{10}$ for $k=2$. The $\mathfrak{gl}_{k|k}$-relative cohomology vanishes through $x^{11}$ for both ranks.
The corresponding characters are therefore $1$ and $0$. This agrees with the integrals
\begin{align}
    Z^+_{0|0}[x;1] &= 1 \, , \qquad Z^+_{1|1}[x;1] = Z^+_{2|2}[x;1] = 0 \, .
\end{align}

\subsection{$\cA=\mathbb{C}[z]$}
\label{subsec:1Cz}

Next we consider one fermionic matrix $\Psi$ and it's derivatives associated to $\cA=\mathbb{C}[z]$ and the single-letter index $f=-x/(1-x)$. 

\paragraph{\underline{$N=0$}} For $N=0$, both relative conditions give
\begin{align}
\chi(H^\bullet(\mathfrak{gl}_{1|1}[\mathbb{C}[z]]|\mathfrak{b}_{1|1})) &= \chi(H^\bullet(\mathfrak{gl}_{1|1}[\mathbb{C}[z]]|\mathfrak{gl}_{1|1})) \\
&=1-x-x^2-x^4-2x^6-x^8-x^9+O(x^{10}) \\
\chi(H^\bullet(\mathfrak{gl}_{2|2}[\mathbb{C}[z]]|\mathfrak{b}_{2|2})) &= \chi(H^\bullet(\mathfrak{gl}_{2|2}[\mathbb{C}[z]]|\mathfrak{gl}_{2|2})) \\
&=1-x-x^2+x^5+x^7+x^9+O(x^{10}).
\end{align}
The integrals $Z^+_{k|k}[f;0]=Z^+_{k|k}[f]$ were computed in \cite{Eniceicu:2023uvd}:
\begin{align}
    Z^{+}_{1|1}[f] &= -x - x^2 - x^4 - 2 x^6 - x^8 - x^9 + \mathcal{O}(x^{10}) \\
    Z^+_{2|2}[f] &= x^4 + x^5 + 2 x^6 + x^7 + x^8 + 2 x^9 + \mathcal{O}(x^{10}) \, .
\end{align}
Their partial sums agree with the graded characters:
\begin{align}
    1+ Z^+_{1|1}[f] &= 1-x - x^2 - x^4 - 2 x^6 - x^8 - x^9 + \mathcal{O}(x^{10}) \\
    1+ Z^+_{1|1}[f]+ Z^+_{2|2}[f] &= 1 - x - x^2 + x^5 + x^7 + x^9 + \mathcal{O}(x^{10}) \, .
\end{align}

\paragraph{\underline{$N=1$}} For $N=1$, the $\mathfrak{b}_{k|k}$-relative characters are:
\begin{align}
\chi(H^\bullet(\mathfrak{gl}_{1|1}[\mathbb{C}[z]]|\mathfrak{b}_{1|1}),\Ber)
&=1-x^2-x^3-x^4-x^6-x^8-x^9+O(x^{10}) \\
\chi(H^\bullet(\mathfrak{gl}_{1|1}[\mathbb{C}[z]]|\mathfrak{b}_{1|1}),\Ber)
&=1-x^2-x^3-x^4+x^7+x^8+x^9+O(x^{10}) \, .
\end{align}
The $\mathfrak{gl}_{k|k}$-relative characters are:
\begin{align}
\chi(H^\bullet(\mathfrak{gl}_{1|1}[\mathbb{C}[z]]|\mathfrak{gl}_{1|1}),\Ber)
&=-x^2-x^3-x^4-x^6-x^8-x^9+O(x^{10}) \\
\chi(H^\bullet(\mathfrak{gl}_{1|1}[\mathbb{C}[z]]|\mathfrak{gl}_{1|1}),\Ber)
&=x^6+x^7+2x^8+2x^9+O(x^{10}) \, .
\end{align}
The integrals $Z^+_{k|k}[f;1]$ were computed in \cite{Eniceicu:2023uvd}:
\begin{align}
    Z^{+}_{1|1}[f;1] &= -x^2 - x^3 - x^4 - x^6 - x^8 - x^9 + \mathcal{O}(x^{10}) \\
    Z^+_{2|2}[f;1] &= x^6 + x^7 + 2 x^8 + 2 x^9 + \mathcal{O}(x^{10}) \, .
\end{align}
Their partial sums are
\begin{align}
    1+ Z^+_{1|1}[f;1] &= 1-x^2 - x^3 - x^4 - x^6 - x^8 - x^9 + \mathcal{O}(x^{10}) \\
    1+ Z^+_{1|1}[f;1] + Z^+_{2|2}[f;1]&= 1 - x^2 - x^3 - x^4 + x^7 + x^8 + x^9 + \mathcal{O}(x^{10}) \, .
\end{align}
We see that the individual integrals match the graded characters of the $\mathfrak{gl}_{k|k}$-relative cohomology, while the partial sums match the $\mathfrak{b}_{k|k}$-relative cohomology.

\section{Large-$k$ Cohomology Conjecture}
\label{app:largek}

In this Appendix, we test the conjecture \eqref{eq:largek}:
\begin{align}
\lim_{k\rightarrow\infty} H^\bullet(\mathfrak{gl}_{k|k}'[\cA^!]|\mathfrak{b}_{k|k},\Ber^N) \simeq \cH_{\cA;N}^\vee \, .
\end{align}
We compute the cohomology at finite $k$ using Python code produced by ChatGPT and Claude, which we intend to include in version 2, with more examples.

\subsection{$\cA=\bC[\theta]$}

First, we consider the one matrix example $\cA=\bC[\theta]$ with $f=x$. The Koszul dual algebra is $\cA^!\simeq\bC[z]$ with $f^!=-x/(1-x)$. The corrections $G^{(k)}_N[f]=Z^+_{k|k}[f^!;N]$ were already considered in Appendix \ref{subsec:1Cz}. The MGGE also includes the large $N$ index:
\begin{align}
    Z_\infty[x] &= \prod_{n=1}^\infty \frac{1}{(1-x^n)} \\
    &= 1 + x + 2 x^2 + 3 x^3 + 5 x^4 + 7 x^5 + 11 x^6 + 15 x^7 + 22 x^8 + 
 30 x^9 + \mathcal{O}(x^{10}) \, .
\end{align}

\paragraph{\underline{$N=0$}} The physical cohomology contains only the identity and $Z_0[x]=1$.

\begin{table}[htbp]
\centering
\begin{tabular}{c|ccc|c}
 & \multicolumn{3}{c|}{$k=1$}
 & \multicolumn{1}{c}{$k=2$} \\

 & $\dim H^0$ & $\dim H^3$ & $\dim H^5$
 & $\dim H^0$ \\
\hline
$x^0$ & 1 &  0 & 0 & 1  \\
$x^1$ & 0 &  0 & 0 & 0  \\
$x^2$ & 0 &  0 & 0 & 0  \\
$x^3$ & 0 &  0 & 0 & 0  \\
$x^4$ & 0 &  1 & 0 & 0  \\
$x^5$ & 0 &  2 & 0 & 0  \\
$x^6$ & 0 &  5 & 0 & 0  \\
$x^7$ & 0 &  8 & 0 & 0  \\
$x^8$ & 0 & 14 & 0 & 0  \\
\end{tabular}
\caption{Dimensions of the cohomology $H^\bullet(\mathfrak{gl}'_{k|k}[\cA^!]|\mathfrak{b}_{k|k})$ for $\cA=\bC[\theta]$. The omitted ghost degrees vanish for the displayed charges.}
\label{tab:theta-N0-stabilization}
\end{table}

The computed finite $k$ cohomology is presented in the Table \ref{tab:theta-N0-stabilization}. We find that:
\begin{itemize}
    \item The ghost number zero cohomology recovers the physical cohomology for both $k=1,2$, which for $N=0$ includes just identity in ghost number zero.
    \item The cohomology at $k=1$ has additional spurious classes at higher ghost numbers, which disappear as we increase to $k=2$.
    \item The graded characters for $k=1,2$ match the partial MGGE sums:
    \begin{align}
    Z_\infty(1+G^{(1)}_0) &= 1 - x^4 - 2 x^5 - 5 x^6 - 8 x^7 - 14 x^8 + \mathcal{O}(x^9) \\
    Z_\infty(1+G^{(1)}_0+G^{(2)}_0) &= 1+ \mathcal{O}(x^9) \, .
    \end{align}
    \item Further checks show that $k=2$ has a spurious class at $x^9$ in $H^8$ which we expect to lift at $k=3$, etc.
\end{itemize}

\paragraph{\underline{$N=1$}} The physical cohomology is $1,X,X^2,\dots$ concentrated in ghost number zero, with character
\begin{align}
    Z_1[x]=\frac{1}{1-x}=1+x+x^2+\dots \, .
\end{align}

\begin{table}[htbp]
\centering
\begin{tabular}{c|cc|c}
 & \multicolumn{2}{c|}{$k=1$} & $k=2$ \\
 & $\dim H^0$ & $\dim H^3$
 & $\dim H^0$ \\
\hline
$x^0$    & 1 &  0 & 1 \\
$x^1$    & 1 &  0 & 1 \\
$x^2$    & 1 &  0 & 1 \\
$x^3$    & 1 &  0 & 1 \\
$x^4$    & 1 &  0 & 1 \\
$x^5$    & 1 &  0 & 1 \\
$x^6$    & 1 &  1 & 1 \\
$x^7$    & 1 &  2 & 1 \\
$x^8$    & 1 &  5 & 1 \\
$x^9$    & 1 &  9 & 1 \\
\end{tabular}
\caption{Dimensions of the cohomology $H^\bullet(\mathfrak{gl}'_{k|k}[\cA^!]|\mathfrak{b}_{k|k},\Ber)$ for $\cA=\bC[\theta]$. The omitted ghost degrees vanish for the displayed charges.}
\label{tab:theta-N1-stabilization}
\end{table}

The computed finite $k$ cohomology is shown in Table \ref{tab:theta-N1-stabilization}. We find that:
\begin{itemize}
    \item The ghost number zero cohomology recovers the physical cohomology for both $k=1,2$.
    \item The cohomology at $k=1$ has additional spurious classes at higher ghost number which disappear as we increase to $k=2$.
    \item The graded characters for $k=1,2$ match the partial MGGE sums:
    \begin{align}
    Z_\infty(1+G^{(1)}_1) &= 1 + x + x^2 + x^3 + x^4 + x^5 \\
    &\qquad - x^7 - 4 x^8 - 8 x^9 + \mathcal{O}(x^{10}) \\
    Z_\infty(1+G^{(1)}_1+G^{(2)}_1) &= 1 + x + x^2 + x^3 + x^4 + x^5 \\
    &\qquad + x^6 + x^7 + x^8 + x^9 + \mathcal{O}(x^{10}) \, .
\end{align}
\end{itemize}

\subsection{$\cA=\bC[z]$}

Next, we consider the example of fermionic matrix with derivatives $\cA=\bC[z]$ with $f=-x/(1-x)$. The Koszul dual algebra is $\cA^!\simeq\bC[\theta]$ with $f^!=x$. The corrections $G^{(k)}_N[f]=Z^+_{k|k}[f^!;N]$ were already considered in Appendix \ref{subsec:1Ctheta}. The MGGE also includes the large $N$ index:
\begin{align}
    Z_\infty\qty[-\frac{x}{1-x}] &=\prod_{n=1}^\infty (1-x^n) = 1 - x - x^2 + x^5 + x^7 + \mathcal{O}(x^{10}) \, .
\end{align}

\paragraph{\underline{$N=0$}} The physical cohomology contains only the identity and $Z_0[x]=1$.

\begin{table}[htbp]
\centering
\begin{tabular}{c|ccc|c}
 & \multicolumn{3}{c|}{$k=1$} & $k=2$ \\
 & $\dim H^0$ & $\dim H^{-1}$ & $\dim H^{-2}$
 & $\dim H^0$ \\
\hline
$x^0$ & 1 & 0 & 0 & 1 \\
$x^1$ & 0 & 0 & 0 & 0 \\
$x^2$ & 0 & 0 & 0 & 0 \\
$x^3$ & 0 & 0 & 0 & 0 \\
$x^4$ & 0 & 1 & 0 & 0 \\
$x^5$ & 0 & 1 & 0 & 0 \\
$x^6$ & 0 & 2 & 0 & 0 \\
$x^7$ & 0 & 2 & 1 & 0 \\
$x^8$ & 0 & 3 & 2 & 0
\end{tabular}
\caption{Dimensions of the cohomology $H^\bullet(\mathfrak{gl}'_{k|k}[\cA^!]|\mathfrak{b}_{k|k})$ for $\cA=\bC[z]$. The omitted ghost degrees vanish for the displayed charges.}
\label{tab:cz-N0-stabilization}
\end{table}

The computed finite $k$ cohomology is presented in the Table \ref{tab:cz-N0-stabilization}. We find that:
\begin{itemize}
    \item The ghost number zero cohomology recovers the physical cohomology for both $k=1,2$, which for $N=0$ includes just identity in ghost number zero.
    \item The cohomology at $k=1$ has additional spurious classes at negative ghost numbers, which disappear as we increase to $k=2$.
    \item The graded characters for $k=1,2$ match the partial MGGE sums:
    \begin{align}
    Z_\infty(1+G^{(1)}_0) &= 1 - x^4 - x^5 - 2 x^6 - x^7 - x^8 + \mathcal{O}(x^9) \\
    Z_\infty(1+G^{(1)}_0+G^{(2)}_0) &= 1 + \mathcal{O}(x^9) \, .
\end{align}
    \item Further checks show that the first spurious $k=2$ class appears at $x^9$ in $H^{-1}$, which we expect to lift at $k=3$.
\end{itemize}

\paragraph{\underline{$N=1$}} In the $\cA=\bC[z]$ case $Z_N=Z_\infty$ for $N>0$, which was recently discussed in \cite{Eleftheriou:2026yct}. The physical cohomology is listed in Table \ref{tab:cz-physical-N1}.

\begin{table}[htbp]
\centering
\begin{tabular}{c|ccccc}
 & $H^0$ & $H^{1}$ & $H^{2}$ & $H^{3}$ & $H^{4}$ \\
\hline
$x^0$    & 1 & 0 & 0 & 0 & 0 \\
$x^1$    & 0 & 1 & 0 & 0 & 0 \\
$x^2$    & 0 & 1 & 0 & 0 & 0 \\
$x^3$    & 0 & 1 & 1 & 0 & 0 \\
$x^4$    & 0 & 1 & 1 & 0 & 0 \\
$x^5$    & 0 & 1 & 2 & 0 & 0 \\
$x^6$    & 0 & 1 & 2 & 1 & 0 \\
$x^7$    & 0 & 1 & 3 & 1 & 0 \\
$x^8$    & 0 & 1 & 3 & 2 & 0 \\
$x^9$    & 0 & 1 & 4 & 3 & 0 \\
$x^{10}$ & 0 & 1 & 4 & 4 & 1
\end{tabular}
\caption{Dimensions of the relative cohomology $\cH_{\cA;N}$ for $\cA=\bC[z]$ and $N=1$.}
\label{tab:cz-physical-N1}
\end{table}

\begin{table}[htbp]
\centering
\begin{tabular}{c|ccccc|ccccc}
 & \multicolumn{5}{c|}{$k=1$}
 & \multicolumn{5}{c}{$k=2$} \\
 & $H^0$ & $H^{-1}$ & $H^{-2}$ & $H^{-3}$ & $H^{-4}$
 & $H^0$ & $H^{-1}$ & $H^{-2}$ & $H^{-3}$ & $H^{-4}$ \\
\hline
$x^0$    & 1 & 0 & 0 & 0 & 0 & 1 & 0 & 0 & 0 & 0 \\
$x^1$    & 0 & 1 & 0 & 0 & 0 & 0 & 1 & 0 & 0 & 0 \\
$x^2$    & 0 & 1 & 0 & 0 & 0 & 0 & 1 & 0 & 0 & 0 \\
$x^3$    & 0 & 1 & 1 & 0 & 0 & 0 & 1 & 1 & 0 & 0 \\
$x^4$    & 0 & 1 & 1 & 0 & 0 & 0 & 1 & 1 & 0 & 0 \\
$x^5$    & 0 & 1 & 2 & 0 & 0 & 0 & 1 & 2 & 0 & 0 \\
$x^6$    & 0 & 1 & 2 & 1 & 0 & 0 & 1 & 2 & 1 & 0 \\
$x^7$    & 0 & 1 & 3 & 1 & 0 & 0 & 1 & 3 & 1 & 0 \\
$x^8$    & 0 & 1 & 3 & 2 & 0 & 0 & 1 & 3 & 2 & 0 \\
$x^9$    & 0 & 1 & 4 & 3 & 0 & 0 & 1 & 4 & 3 & 0 \\
$x^{10}$ & 0 & 1 & 4 & 4 & 1 & 0 & 1 & 4 & 4 & 1
\end{tabular}
\caption{Dimensions of the cohomology $H^\bullet(\mathfrak{gl}'_{k|k}[\cA^!]|\mathfrak{b}_{k|k},\Ber)$ for $\cA=\bC[z]$. The omitted ghost degrees vanish for the displayed charges.}
\label{tab:cz-N1-stabilization}
\end{table}

The finite $k$ cohomology is shown in Table \ref{tab:cz-N1-stabilization}. We observe that:
\begin{itemize}
    \item For both $k=1,2$, the computed cohomology agrees with the (dual) of the physical cohomology from Table \ref{tab:cz-physical-N1}. In particular, there is no spurious classes up to the degrees studied.
    \item The corrections $G^{(k)}_1[-x/(1-x)]=Z^+_{k|k}[x;1]$ are zero for $k>0$ and $N>0$, which implies $Z_N=Z_\infty$. Therefore the graded characters match
\begin{align}
Z_1=Z_\infty= 1 - x - x^2 + x^5 + x^7 + \mathcal{O}(x^{11}) \, .
\end{align}
\end{itemize}

\bibliographystyle{JHEP}

\bibliography{mono}

\end{document}